\documentclass[trackchanges, twocolumn]{aastex701}
\usepackage{amsmath}
\usepackage{xcolor}
\begin{document}

\title{Constraints on the $r$-process Delay Time from Dwarf Spheroidal Galaxies}

\author[orcid=0000-0001-6207-4388,sname='Henderson', gname='Lauren']{Lauren E. Henderson}
\affiliation{Department of Physics and Astronomy, University of Notre Dame, 225 Nieuwland Science Hall, Notre Dame, IN 46556, USA}
\email[show]{lhender6@nd.edu} 

\author[orcid=0000-0001-6196-5162,sname='Kirby', gname='Evan']{Evan N. Kirby}
\affiliation{Department of Physics and Astronomy, University of Notre Dame, 225 Nieuwland Science Hall, Notre Dame, IN 46556, USA}
\email{ekirby@nd.edu} 

\author[orcid=0000-0002-4739-046X,sname='de los Reyes', gname='Mithi']{Mithi A. C. de los Reyes}
\affiliation{Department of Physics and Astronomy, Amherst College, 6 East Drive, Amherst, MA 01002, USA}
\email{}

\author[orcid=0000-0002-4729-8823,sname='Surman', gname='Rebecca']{Rebecca Surman}
\affiliation{Department of Physics and Astronomy, University of Notre Dame, 225 Nieuwland Science Hall, Notre Dame, IN 46556, USA}
\email{}

\begin{abstract}

The ancient stellar populations and simple structure of the local dwarf spheroidal galaxies (dSphs) make them excellent laboratories for studying the chemical evolution of the early universe. In this paper, we fit one-zone galactic chemical evolution models to C, Mg, Si, Ca, Fe, Sr, Y, Ba, and Eu abundances in the dSphs Sculptor and Draco. By including the $s$- and $r$-processes in our model, we are able to place constraints on the timescale of $r$-process enrichment in these dSphs. We find that neutron star mergers cannot be the only source of the $r$-process in Sculptor and Draco, and that a prompt $r$-process source is required to explain the observed stellar abundances. When neutron star mergers are the only source of the $r$-process in our models, their minimum delay time must be just 14-16 Myr. The neutron-capture abundances in Sculptor and Draco can also be fit by models with a longer neutron star merger minimum delay time combined with a prompt $r$-process source that operates on the same timescale as massive star evolution, such as magneto-rotational supernovae. Additionally, the Sr and Y abundances in the dSphs set limits on the Sr and Y yields per $r$-process event. We find that theoretical $r$-process yields overestimate Sr and Y production by a factor of up to 100.

\end{abstract}



\section{Introduction}
Old stellar populations in the Local Group provide a window into nucleosynthesis and galaxy evolution in the early universe that is not yet accessible with high-redshift observations. Old stars lock in the composition of the gas from which they were formed, essentially becoming a fossil-record of ancient nucleosynthesis. Among such stellar populations, dwarf spheroidal galaxies (dSphs) are ideal for studying galaxy evolution because they are large enough to have hundreds to tens of thousands of bright stars, allowing for studies at a population-level, but they are small enough to have relatively simple chemical evolution histories. This simplicity, as opposed to larger galaxies like the Milky Way, means that they can be reasonably approximated by analytic chemical evolution models \citep[e.g.,][]{pagel_1997, carigi_dSphGCE_2002, kirby_multi-element_alpha_2011, homma_chemical_2015, kvasova_GCEmodel_2025}. We make use of such models in this paper by using dSphs to study the rapid neutron-capture process ($r$-process) in the early universe.

\subsection{The $r$-process}
Nuclei heavier than iron cannot be created in hydrostatic burning in stars; instead they are almost entirely produced by neutron-capture processes. The slow neutron-capture process ($s$-process) and the rapid one ($r$-process) are each responsible for about half of the heavy elements. While the $r$-process was theorized decades ago \citep{burbidge_synthesis_1957}, its astrophysical sites are still uncertain.

The $r$-process requires a high neutron fluence. The gravitational wave observations and electromagnetic follow-up of the neutron star merger (NSM) GW170817 suggest that NSMs are sites of the $r$-process \citep{coulter_swope_2017, drout_light_2017, kasen_origin_2017, kilpatrick_electromagnetic_2017, pian_NSM_2017, smartt_kilonova_2017}, but they may not be responsible for all of the $r$-process enrichment in the universe. Evidence from metal-poor stars in the Milky Way and other Local Group galaxies suggest that $r$-process elements were created in the very early universe, before NSMs would be able to occur \citep{argast_r-process_2004, cescutti_NSMsMWHalo_2015, ji_r-process_2016, skuladottir_neutron-capture_2019, cain_rIII_2020, molero_evolution_2021, bandyopadhyay_rII_2024, shah_rstar_2024, molero_MWrprocess_2026}. While the minimum time it takes a NSM to occur after star formation, known as the minimum delay time, is not well constrained, it is unlikely that neutron stars would be able to inspiral quickly enough to produce all of the $r$-process abundances observed in old stars \citep{chattaraj_nsmdelaytime_2026}.

There are proposed sites of the $r$-process other than NSMs that could account for early $r$-process enrichment. Magneto-rotational supernovae \citep[MRSNe,][]{winteler_MRSN_2012, nishimura_r-process_2015, nishimura_MRSN_2017, reichert_MRSNe_2021}, electron-capture supernovae \citep{wanajo_ECSN_2011}, proto-neutron star winds \citep{meyer_proto_NS_1992, thompson_protoNS_2001, prasanna_protomagnetar_2025}, collapsars \citep{mclaughlin_surman_2005_collapsars,siegel_2019_collapsars,miller_2020_collapsars}, magnetar flares \citep{cehula_magnetarflares_2024}, and others \citep{fuller_2017_PBH,pitik_2026_AIC} have been proposed, but none have been observationally confirmed. While the studies of the $r$-process in the Milky Way and the Local Group demonstrate the need for rapid $r$-process enrichment, dSphs provide an opportunity to study the timescale of the $r$-process in ancient stellar populations in detail.

\subsection{Galactic chemical evolution}
Galactic chemical evolution (GCE) models track the star formation, gas inflows/outflows, and nucleosynthesis that govern the build-up of metals in a galaxy. They are typically designed to, at minimum, fit iron-group elements (V, Cr, Mn, Fe, Co, and Ni), which are primarily produced in Type Ia supernovae (SNeIa), and $\alpha$ elements (O, Ne, Mg, Si, S, Ar, Ca, and Ti), which are primarily produced in core-collapse supernovae (CCSNe) \citep[e.g.,][]{kobayashi_CtoZn_2006, kirby_multi-element_alpha_2011, vincenzo_chemical_2014, weinberg_equilibrium_2017, hasselquist_apogee_2021, alexander_SNIaGCE_2025}. The $\alpha$ elements thus closely follow the star formation of the galaxy while iron-group elements from SNeIa are more delayed. These two groups of elements are prioritized in GCE models because they are abundant and Mg, Si, Ca, Ti, and Fe are among the easiest elements to measure spectroscopically.

Many GCE models in the literature include the $s$-process and the $r$-process as well. The $s$-process provides another constraint on timescales in the models because it occurs primarily in asymptotic giant branch (AGB) stars, meaning the $s$-process delay time is relatively well-understood \citep{gallino_AGBs_1998, lugaro_s-process_2003, karakas_slar_2016, karakas_SMCAGBs_2018}. The $r$-process is valuable to include in GCE models because doing so can lend clues as to its origin. 

Inclusion of the $s$- and $r$-processes in GCE models is difficult because one must disentangle the contribution of each process to a particular element. Elements such as Sr, Y, and Ba that are often attributed to the $s$-process are also produced by the $r$-process. An element produced almost entirely by the $r$-process, such as Eu, is needed to distinguish the various neutron-capture contributions to other elements \citep{simmerer_rise_2004}. The challenge is that a sufficiently large number of abundance measurements are necessary for this task, and the neutron-capture elements can be difficult to measure spectroscopically because many of their absorption lines are relatively weak.

Neutron-capture processes have been incorporated into GCE models for the Milky Way \citep{matteucci_MWEu_2014, kobayashi_origin_2020, horta_neutron_2022, molero_MWrprocess_2026}, but interpretation of the results is complicated by the Milky Way's complex star formation and merger history. Even so, these studies have demonstrated a need for an $r$-process source with a short delay time.

The earliest study of neutron-capture elements in dSphs was \citet{shetrone_draco_1998}, who found that Ba in the Draco dSph was formed more promptly than Fe.  \citet{letarte_high-resolution_2010} found clear evidence for $s$-process enrichment of Fornax from measurements of Y and Ba.  \citet{duggan_neutron_2018} and \citet{skuladottir_neutron-capture_2019} arrived at different conclusions about the delay time of the $r$-process in applying GCE models of neutron-capture elements to Sculptor. \citet{molero_evolution_2021} fit GCE models to Ba and Eu abundances in nearby dwarf galaxies measured by \citet{reichert_neutron-capture_2020}, and \citet{palla_Eu_2025} focused on fitting Eu abundances from dwarf galaxies and the Milky Way simultaneously. Both of these studies made progress toward using dSphs to infer the timescale of $r$-process production, but were not able to benefit from inclusion of neutron-capture elements besides Ba and Eu. Using medium-resolution spectroscopy with Keck/DEIMOS, \citet{henderson_491_2025} measured Sr, Y, Ba, and Eu abundances for hundreds of stars in local dSphs, providing the sample used in this paper. 
This sample provides $s$- and $r$-process element abundances at a wide range of metallicities in Sculptor and Draco, making it possible for us to perform detailed analysis of the timescale of neutron-capture enrichment in the early universe.




This paper is structured as follows. In Section \ref{sec:methods} we describe our GCE model and our method for fitting the model to dSph stellar abundances. Section \ref{sec:results} details our tests with various SNIa delay times and the best-fit models for different $r$-process sources. We discuss the implications of those tests for the $r$-process in the early universe in Section \ref{sec:discussion} and conclude in Section \ref{sec:conc}.

\section{Methods} \label{sec:methods}
\subsection{Abundance measurements} \label{subsec:abundances}

We collected abundances from the literature for red giant stars in Sculptor and Draco: [Fe/H], [Si/Fe], and [Ca/Fe] from \citet{kirby_multi-element_2010}; [C/Fe] from \citet{kirby_carbon_2015} (corrected for stellar evolution following \citealt{placco_carbon-enhanced_2014}); and [Mg/Fe], [Sr/Fe], [Y/Fe], [Ba/Fe], and [Eu/Fe] from \citet{henderson_491_2025}. All of these abundances are from spectra obtained with the DEep Imaging Multi-Object Spectrograph (DEIMOS) on the Keck II telescope \citep{faber_deimos_2003}. 



\subsection{The chemical evolution model} \label{subsec:GCEmodel}

The GCE model we used in this study is an open-box, one-zone model with instantaneous mixing. This type of model makes many simplifying assumptions:
\begin{enumerate}
    \item one-zone: all of the gas is assumed to be well-mixed at all times. This prevents us from being able to represent scatter at a given metallicity or spatially in our models.
    \item instantaneous mixing: the metals produced by supernovae, AGB stars, and $r$-process events are immediately mixed with all of the gas in the galaxy.
    \item parameterized quantities: as we will describe in more detail below, the star formation, inflows, and outflows are all parameterized in some way. This may limit how well our models can describe specific dwarf galaxies if their evolution is not well-described by the parameterization.
\end{enumerate}
In general, these assumptions are reasonable for dSphs because of their small size and simple structure \footnote{This is not true for Fornax, which is why it is not included in the main portion of this paper. Discussion of Fornax can be found in Appendix \ref{appendix:fornax}.}. For more on the limitations of these models, see Section \ref{subsec:error}. We implemented a modified version of the model presented by \citet[][dlR22 hereafter]{delosreyes_simultaneous_2022} and \citet{kirby_multi-element_alpha_2011}. For clarity, we summarize the key aspects here.

The general form of the GCE model is presented in Equation \ref{eq:GCE}, where $\xi_{j}(t)$ is the gas-phase abundance of element $j$. Ejecta from nucleosynthesis events enrich the gas with metals, while star formation and outflows deplete the gas and inflowing gas dilutes the metals. The gas-phase abundances of metals are tracked over discrete time steps of 1 Myr. The biggest difference between this model and that used in dlR22 is the tracking of $r$-process elements and the inclusion of $r$-process sources, like NSMs and MRSNe ($\dot{\xi}_{j,r}$ in Equation \ref{eq:GCE}).
\begin{equation} \label{eq:GCE}
\begin{split}
    \xi_{j}(t) = & \int_{0}^{t}(\dot{\xi}_{j,{\rm II}} + \dot{\xi}_{j,{\rm AGB}} + \dot{\xi}_{j,{\rm Ia}} + \dot{\xi}_{j,r} - \dot{\xi}_{j,{\rm SF}}\\
     & + \dot{\xi}_{j,{\rm in}} - \dot{\xi}_{j,{\rm out}})dt
\end{split}
\end{equation}

Star formation is parameterized by the Kennicutt-Schmidt law \citep{schmidt_SF_1959, kennicutt_sflaw_1998}:
\begin{equation} \label{eq:star_formation}
    \dot{\xi}_{j,{\rm SF}}(t) = A_{*}\left(\frac{M_{j,{\rm gas}}(t)}{10^6~M_{\odot}}\right)^{\alpha}
\end{equation}
where the star formation rate is directly related to the gas mass in the galaxy. $A_*$ and $\alpha$ are treated as free parameters. 

The inflowing gas has a pristine composition of H and He from Big Bang nucleosynthesis: mass fractions of 0.7514 and 0.2486, respectively \citep{cyburt_bigbang_2008}. The rate of inflow is given by:
\begin{equation} \label{eq:inflow}
    \dot{\xi}_{j,{\rm in}}(t) = A_{\rm in}\frac{M_{j,{\rm gas}}(0)}{M_{\rm gas}(0)}\left(\frac{t}{\rm Gyr}\right)e^{-t/\tau_{\rm in}}
\end{equation}
where $A_{\rm in}$ controls the gas inflow rate and $\tau_{\rm in}$ represents the gas inflow timescale. This parameterization of the gas inflow corresponds to a fast increase in inflow followed by a slow decrease and is commonly used in dSph chemical evolution models \citep{lynden-bell_GCE_1975, kirby_multi-element_alpha_2011} for its ability to reproduce observed MDFs. Chemical evolution models also need outflows to describe dSphs. These outflows are powered by supernovae, and thus the outflow rate is proportional to the supernova rate:
\begin{equation} \label{eq:outflows}
    \dot{\xi}_{j,{\rm out}}(t) = A_{\rm out}\frac{M_{j,{\rm gas}}(t)}{M_{\rm gas}(t)}(\dot{N}_{\rm II} + \dot{N}_{\rm Ia}) .
\end{equation}
These outflows begin with the onset of the first SNe in the galaxy; no other physical constraints or fine-tuning are applied. We assume the processes producing $r$-process elements, such as neutron star mergers, are rare enough that CCSNe and SNeIa dominate the outflow rate.

The equations described above determine the chemically deduced star formation history (SFH) of the galaxy. We did not to impose a photometric SFH from the literature, instead allowing the fit to the stellar abundances to determine the SFH\@. As noted by dlR22, independent spectroscopic SFHs are a useful complement to photometric SFHs for old dSphs. Model isochrones have poor age resolution at old stellar ages and low metallicities, making it difficult to constrain the SFHs of galaxies like Sculptor and Draco from photometry alone. The chemical evolution, on the other hand, is sensitive to the duration of star formation in these ancient galaxies.

\subsection{Delay time distributions} \label{subsec:DTD}

$\dot{\xi}_{j,{\rm II}}$, $\dot{\xi}_{j,{\rm AGB}}$, $\dot{\xi}_{j,{\rm Ia}}$, and $\dot{\xi}_{j,r}$ at a particular time step are determined by the prior star formation and the delay time distribution (DTD) for each type of nucleosynthesis event. The delay time distribution describes how long after a burst of star formation each event is likely to occur. As in dlR22, our GCE model computes how many nucleosynthesis events will happen in the future and propagates the future time steps of an array with elemental abundances. 

For CCSNe and AGB stars, the DTD is governed by stellar evolution. The number of CCSNe and AGB stars depends on the initial mass function (IMF), for which we use the \citet{kroupa_IMF_1993} IMF\@. We assume that all stars born with 10--100 $M_{\odot}$ explode as CCSNe at the end of their lives. Likewise, we assume that all stars $<$10 $M_{\odot}$ will evolve onto the AGB and expel their mass as winds over the last 1 Myr of their lives \citep{marigo_AGB_2007}. The stellar lifetimes as a function of mass are parameterized following \citet{padovani_massloss_1993} (Equations (8) and (9) in dlR22).

Our model has the option of including a prompt $r$-process source that occurs on the same timescale as CCSNe. We choose MRSNe, but other hypothesized prompt $r$-process sources could fill this roll as well. We set a fraction of CCSNe of stars with initial masses between 10 and 35 $M_{\odot}$ to be MRSNe and produce $r$-process elements. This fraction ($f_{\rm MRSN}$) is allowed to be a free parameter in the model.

The DTDs for SNeIa and NSMs are more complicated than prompt sources. SNeIa can occur at a range of times, with the minimum delay time roughly corresponding to at least the time it takes the most massive SNIa progenitor star to become a white dwarf. The most accurate parameterization for the SNIa DTD is still an open question (see \citealt{dubay_extended_2024, heiger_dleiy_2026} for recent work on the topic). We parameterize the SNIa DTD with a power law \citep{maoz_IaDTD_2010}:
\begin{equation} \label{eq:IaDTD}
    \Psi_{\rm Ia} = 
    \begin{cases}
        0 & \text{if } t < t_{\rm min, Ia}\\
        (A_{\rm Ia}~{\rm Gyr}^{-1}~M^{-1}_{\odot})\left(\frac{t}{\rm Gyr}\right)^{-1.1} & \text{if } t \geq t_{\rm min, Ia}
    \end{cases}
\end{equation}
The minimum SNIa delay time, $t_{\rm min}$, and the normalization, $A_{\rm Ia}$, can be free parameters in the model. We discuss the SNIa DTD in more detail in Section \ref{subsec:Ia_discussion}.

In addition to MRSNe, NSMs are also a source of the $r$-process in our GCE model. Like SNeIa, the NSM DTD is an open question. We parameterize the NSM DTD as:
\begin{equation} \label{eq:NSMDTD}
    \Psi_{\rm NSM} = 
    \begin{cases}
        0 & \text{if } t < t_{\rm min, NSM}\\
        A_{\rm NSM}(10^{-4}~{\rm Gyr}^{-1}~M^{-1}_{\odot})\left(\frac{t}{\rm Gyr}\right)^{-1.5} & \text{if } t \geq t_{\rm min, NSM}
    \end{cases}
\end{equation}
following \citet{simonetti_nsmDTD_2019}. We allow the normalization, $A_{\rm NSM}$, and the minimum delay time, $t_{\rm min}$, to be free parameters.

\subsection{Yields} \label{subsec:yields}
In addition to the basic equations and assumptions that describe chemical evolution, another key input for our model is the nucleosynthesis yields. Theoretical yield sets vary due to limited constraints on reaction rates, uncertain or difficult-to-model physics, and uncertainty in progenitors. For CCSNe and SNeIa, we use the same yields as dlR22. These are based on \citet{nomoto_stellarnuc_2013} and \citet{limongichieffi_2018} for CCSNe and on \citet{leung_MCh_2018}, \citet{shen_subMCh_2018}, and \citet{leung_subMCh_2020} for SNeIa. We also use the best-fit normalizations (yield per event) they found for Fe, Mg, etc.

\begin{figure}
    \centering
    \includegraphics[width=1.0\linewidth]{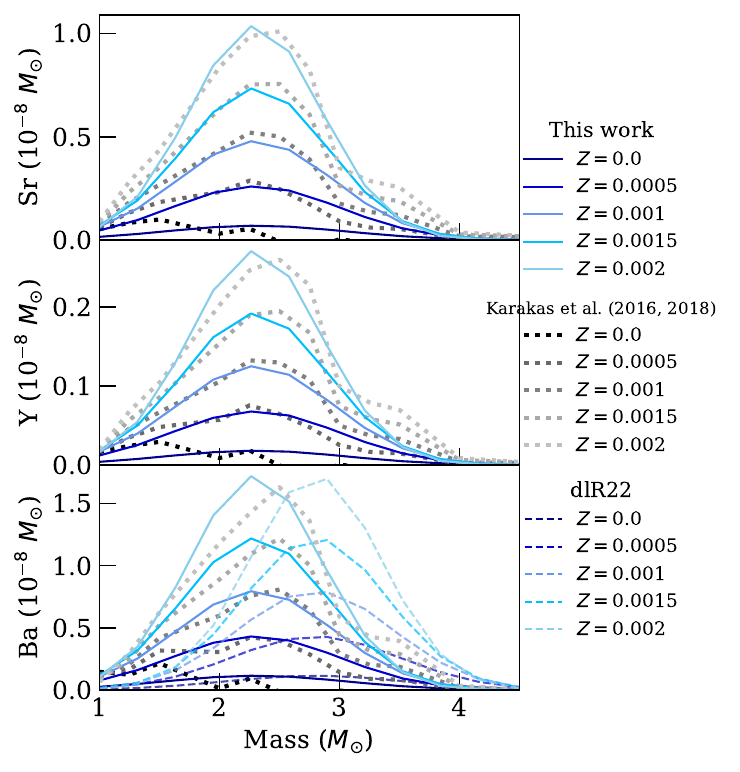}
    \caption{Sr, Y, and Ba yields from AGB stars as a function of initial stellar mass. The yields are weighted by the \citet{kroupa_IMF_1993} IMF. Different colors represent different metallicities of the progenitor star. The solid lines show the yields used in this work, with the gray dotted lines showing the yields of \citet{karakas_slar_2016} and \citet{karakas_SMCAGBs_2018} for comparison. In the bottom panel, the dashed blue lines show the best-fit Ba yields used in dlR22.}
    \label{fig:AGByields}
\end{figure}

The AGB yields used in dlR22 that we base our yields on are from \citet{cristallo_fruity_2015}, \citet{karakas_slar_2016}, and \citet{karakas_SMCAGBs_2018}.
The AGB yields in dlR22 were parameterized by an equation that approximates the theoretical yield as a function of stellar mass to allow for more flexibility when editing the yields. Our AGB yields differ slightly from those in dlR22. Instead of using an empirical mass peak for Ba at 2.80~$M_{\odot}$, we shift it down to 2.3 $M_{\odot}$ to better match the \citet{karakas_slar_2016} and \citet{karakas_SMCAGBs_2018} yields. We show our Sr, Y, and Ba AGB yields in Figure \ref{fig:AGByields}. The Ba yields used in dlR22 are shown in the bottom panel for comparison.

The NSM yields in this study are from \citet{holmbeck_MWNSMs_2024}. These yields are based on the binary neutron star population of the Milky Way. We found that these yields work well for [Ba/Eu], but over-produce weak $r$-process elements (Sr and Y) in comparison to Ba and Eu. As we will describe in more detail in Section \ref{subsec:SrandYnorms}, we apply a normalization to the Sr and Y NSM yields.

We also included MRSNe in this model. We adapted $r$-process yields from \citet{reichert_MRSNe_2021} (the 35OC-Rs model), who calculated the yields from the death of a 35 $M_{\odot}$ star. Like for NSMs, we added a normalization term to the Sr and Y yields. The Sr and Y normalization terms for both NSMs and MRSNe can be found in Table \ref{tab:r_tests}.

\subsection{Fitting the models to data} \label{subsec:fitting}
We fit the GCE model to spectroscopic stellar abundances in the dSphs (see Section \ref{subsec:abundances}): [Fe/H], [C/Fe], [Mg/Fe], [Si/Fe], [Ca/Fe], [Sr/Fe], [Y/Fe], [Ba/Fe], and [Eu/Fe]. Our method for fitting the model to the data simultaneously takes into account the model's trajectory through the 9-dimensional space created by each element and the metallicity distribution function (MDF; how many stars are found at each [Fe/H] value).

Our likelihood function is structured in the same way as in dlR22, with the addition of upper limits. Including upper limits in the likelihood is important in our case because the neutron-capture abundances many more upper limits than the other elements, especially at low metallicity. In order to avoid overestimating the neutron-capture GCE trends at early times, we need to account for these upper limits.

The contribution to the likelihood function for a single star, $L_i$ (Equation \ref{eq:starlikelihood}), therefore has two possible components for each element: a probability distribution function if the element has a measured abundance, and a cumulative distribution function if the element only has an upper limit estimate (Equation \ref{eq:li_upper_or_not}).

\begin{equation} \label{eq:starlikelihood}
    L_i = \int_{0}^{t_{final}}\prod_j\frac{1}{\sqrt{2\pi}\sigma_{ij}}\ell_{ij}\frac{\dot{M}_*(t)}{M_*}dt
\end{equation}

\begin{equation} \label{eq:li_upper_or_not}
    \ell_{ij} = 
    \begin{cases}
        \exp\left(-\frac{(\epsilon_{ij} - \epsilon_j(t))^2}{2\sigma_{ij}^2}\right), & \text{if not an upper limit} \\
        \int_{-\infty}^{\epsilon_{ij}}\exp\left(-\frac{(u - \epsilon_j(t))^2}{2\sigma_{ij}^2}\right)du, & \text{if an upper limit}
    \end{cases}
\end{equation}

In Equations \ref{eq:starlikelihood} and \ref{eq:li_upper_or_not} $\epsilon_{ij}$ is the measured abundance (or upper limit) for element $j$ in star $i$; $\sigma_{ij}$ is the corresponding error on the measurement. $\epsilon_j(t)$ is the [X/Fe] abundance of element $j$ as a function of time in the model.

The final likelihood for the model is calculated by combining $L_i$ for each of the $N$ stars and adding a term that represents how appropriate the model is to the galaxy's overall properties (Equation \ref{eq:finalL}; this is the same as Equation 11 of dlR22). In particular, we enforce that the final stellar and gas masses of the model be reasonable matches to the galaxy's observed properties. We set the final gas mass to zero for all the galaxies in our sample with $\delta M_{\rm gas,{\rm obs}} = 1000$. We use stellar masses and errors from \citet{woo_dwarfscaling_2008}. The exponent $0.1N$ in Equation \ref{eq:finalL} weights the last term relative to the stellar abundance contribution to ensure that this term does not dominate the likelihood.  It is effectively a prior that reflects our belief that the stellar abundances are more trustworthy representations of chemical evolution than the final gas mass.

\begin{equation} \label{eq:finalL}
\begin{split}
    L = &\prod_i^NL_i\times\left(\frac{1}{\sqrt{2\pi}\delta M_{*,{\rm obs}}}\exp\left(\frac{-(M_{*,{\rm obs}}-M_{*,{\rm model}})^2}{2(\delta M_{*,{\rm obs}})^2}\right) \right. \\
    & \left. \times\frac{1}{\sqrt{2\pi}\delta M_{\rm gas,{\rm obs}}}\exp\left(\frac{-(M_{\rm gas,{\rm obs}}-M_{\rm gas,model})^2}{2(\delta M_{\rm gas,obs})^2}\right)\right)^{0.1N}
\end{split}
\end{equation}

We use \texttt{emcee}, a Markov Chain Monte Carlo (MCMC) Python package, to fit the GCE model to the data \citep{emcee_2013}. Our GCE model is flexible; we can choose any combination of the free parameters given in Table \ref{tab:params} to fit the data, given that $A_{\rm in}$, $\tau_{\rm in}$, $A_{\rm out}$, $A_*$, and $\alpha$ are included. These parameters must be fit in each model because they govern the gas mass and star formation in the galaxy. Each free parameter has flat, uniform priors, listed in Table \ref{tab:params}. These priors simply enforce physically-meaningful values for the parameters. The initial guess for each parameter is either the best-fit value for Sculptor from dlR22 or a reasonable value, in the case of nucleosynthesis parameters not included in dlR22. We run each \texttt{emcee} calculation with 32 walkers and 15625 steps, discarding the first 5000 steps as ``burn-in.'' The remaining steps then sample the posterior distribution of the free parameters.

Some walkers got stuck at local minima in the parameter space that were not good fits to the observations. These walkers were therefore not able to explore the parameter space and do not provide information for determining best-fit parameters. We filtered out walkers that ended their chain at a $-lnL$ value less than 3 times the interquartile range to remove extreme outliers. The best-fit values and uncertainty intervals in Tables \ref{tab:Iatests} and \ref{tab:r_tests} were calculated after the spurious walkers were removed.

\begin{deluxetable}{cccc}
\tablecolumns{4}
\tablewidth{0pt}
\tablecaption{Model Parameters\label{tab:params}}
\tablehead{
 \colhead{Parameter} & \colhead{Description} & \colhead{Prior} & \colhead{Initial Guess}}
\startdata
$A_{\rm in}$ & Normalization of gas inflow rate ($10^9~M_{\odot}~{\rm Gyr}^{-1}$) & $\mathcal{U}(0,5)$ & 0.53 \\
$\tau_{\rm in}$ & Gas inflow time constant (Gyr) & $\mathcal{U}(0,1)$ & 0.27 \\
$A_{\rm out}$ & Gas lost per supernova ($M_{\odot}~{\rm SN}^{-1}$) & $\mathcal{U}(0,20)$ & 4.79 \\
$A_{*}$ & Normalization of star formation law ($10^6~M_{\odot}~{\rm Gyr}^{-1}$) & $\mathcal{U}(0,10)$ & 0.79 \\
$\alpha$ & Power-law index of star formation law & $\mathcal{U}(0,2)$ & 0.72 \\
$A_{\rm Ia}$ & Normalization of SNIa DTD ($10^{-3}~M_{\odot}~{\rm Gyr}^{-1}$) & $\mathcal{U}(0,10)$ & 1.0 \\
$t_{\rm min,Ia}$ & Minimum SNIa delay time (Gyr) & $\mathcal{U}(0,10)$ & 0.10 \\
$A_{\rm NSM}$ & Normalization of NSM DTD ($10^{-4}~M_{\odot}~{\rm Gyr}^{-1}$) & $\mathcal{U}(0.5,10)$ & 3.5 \\
$t_{\rm min,NSM}$ & Minimum NSM delay time (Gyr) & $\mathcal{U}(0.01,0.1)$ & 0.01 \\
${\rm normSr}_{\rm NSM}$ & Normalization of NSM Sr yield & $\mathcal{U}(0,1)$ & 1.0 \\
${\rm normY}_{\rm NSM}$ & Normalization of NSM Y yield & $\mathcal{U}(0,1)$ & 1.0 \\
$f_{\rm MRSN}$ & Fraction of 10--35 $M_{\odot}$ stars that explode as MRSNe & $\mathcal{U}(0,1)$ & 0.1 \\
${\rm normSr}_{\rm MRSN}$ & Normalization of MRSN Sr yield & $\mathcal{U}(0,1)$ & 1.0 \\
${\rm normY}_{\rm MRSN}$ & Normalization of MRSN Y yield & $\mathcal{U}(0,1)$ & 1.0 \\
\enddata
\end{deluxetable}

\section{Applying the GCE model} \label{sec:results}

\begin{figure*}
    \centering
    \includegraphics[width=1.0\linewidth]{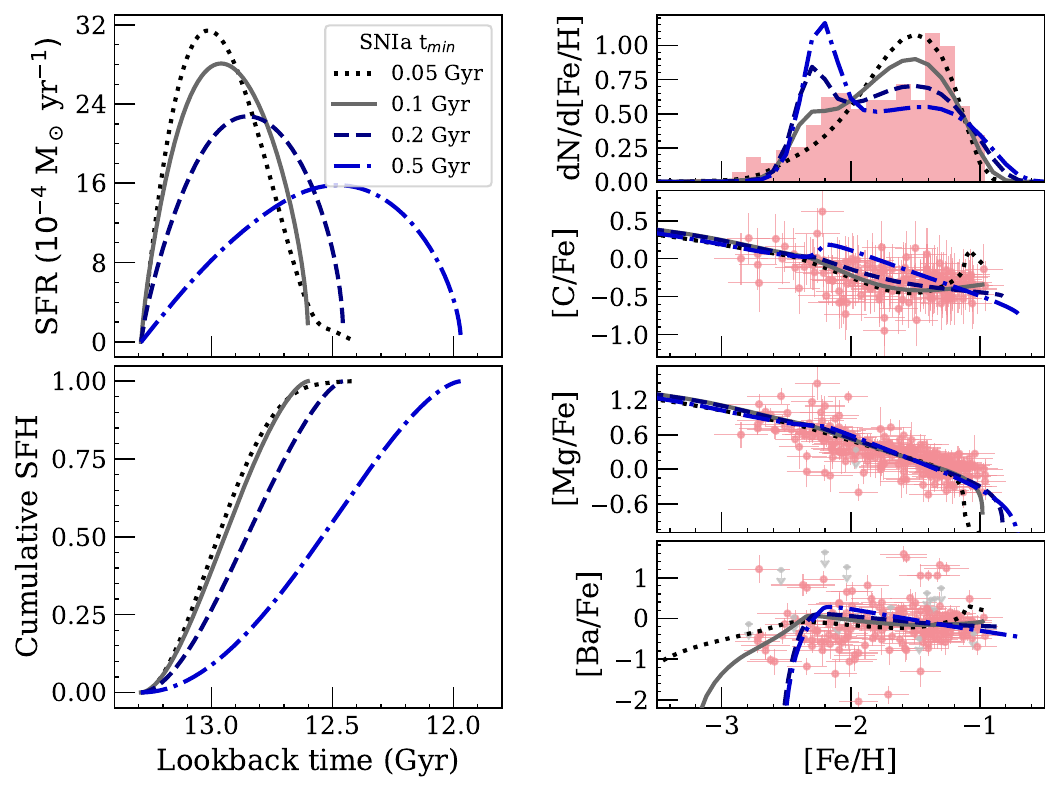}
    \caption{Left: star formation rate over time (top) and cumulative star formation history (bottom) for 4 different GCE models for Sculptor. Each model has a different SNIa minimum delay time, $t_{\rm min,Ia}$, from 0.05 Gyr to 0.5 Gyr. Right: Metallicity distribution function (top) and [C/Fe], [Mg/Fe], and [Ba/Fe] as a function of [Fe/H] for Sculptor. Pink points are measurements, gray arrows are upper limits. Plotted over the measurements are the same 4 GCE models from the left panels. The models were fit to all the elements in our sample; C, Mg, and Ba are shown as a representative set of elements. Model parameters are given in Table \ref{tab:Iatests}.}
    \label{fig:Scl_Iatmin}
\end{figure*}

\subsection{Type Ia supernova tests} \label{subsec:Iatests}
The timescale of Fe enrichment in the galaxies is largely determined by the minimum delay time and DTD of Type Ia SNe. Thus, the DTD chosen for SNeIa can significantly affect the timescales we infer from our models, such as the minimum delay time for NSMs and the SFHs of the galaxies. 

We ran a suite of model optimizations for Sculptor to test whether the models and data are sensitive to the minimum delay time of SNeIa, and whether that will significantly change our conclusions. The tests are summarized in Table \ref{tab:Iatests}. In each test, the parameterization of the SNIa DTD remained the same; only $t_{\rm min, Ia}$ and $A_{\rm Ia}$ changed (Equation \ref{eq:IaDTD}). We fit GCE models to the Sculptor abundances with 4 different SNIa minimum delay times: 50, 100, 200, and 500 Myr. In each MCMC run $A_{\rm in}$, $\tau_{\rm in}$, $A_{\rm out}$, $A_*$, $\alpha$, $A_{\rm Ia}$, $t_{\rm min,NSM}$, $A_{\rm NSM}$, ${\rm normSr}_{\rm NSM}$, and ${\rm normY}_{\rm NSM}$ were free parameters.

Figure \ref{fig:Scl_Iatmin} compares the best-fit models for each of the 4 SNIa minimum delay times in our test. The left panels show the effect of the SNIa delay time on the SFH timescale. In the right panels, [C/Fe], [Mg/Fe], and [Ba/Fe] are shown as representative elements. In Section \ref{subsec:Ia_discussion} we discuss the implications of these tests in detail.

\begin{figure}
    \centering
    \includegraphics[width=1.0\linewidth]{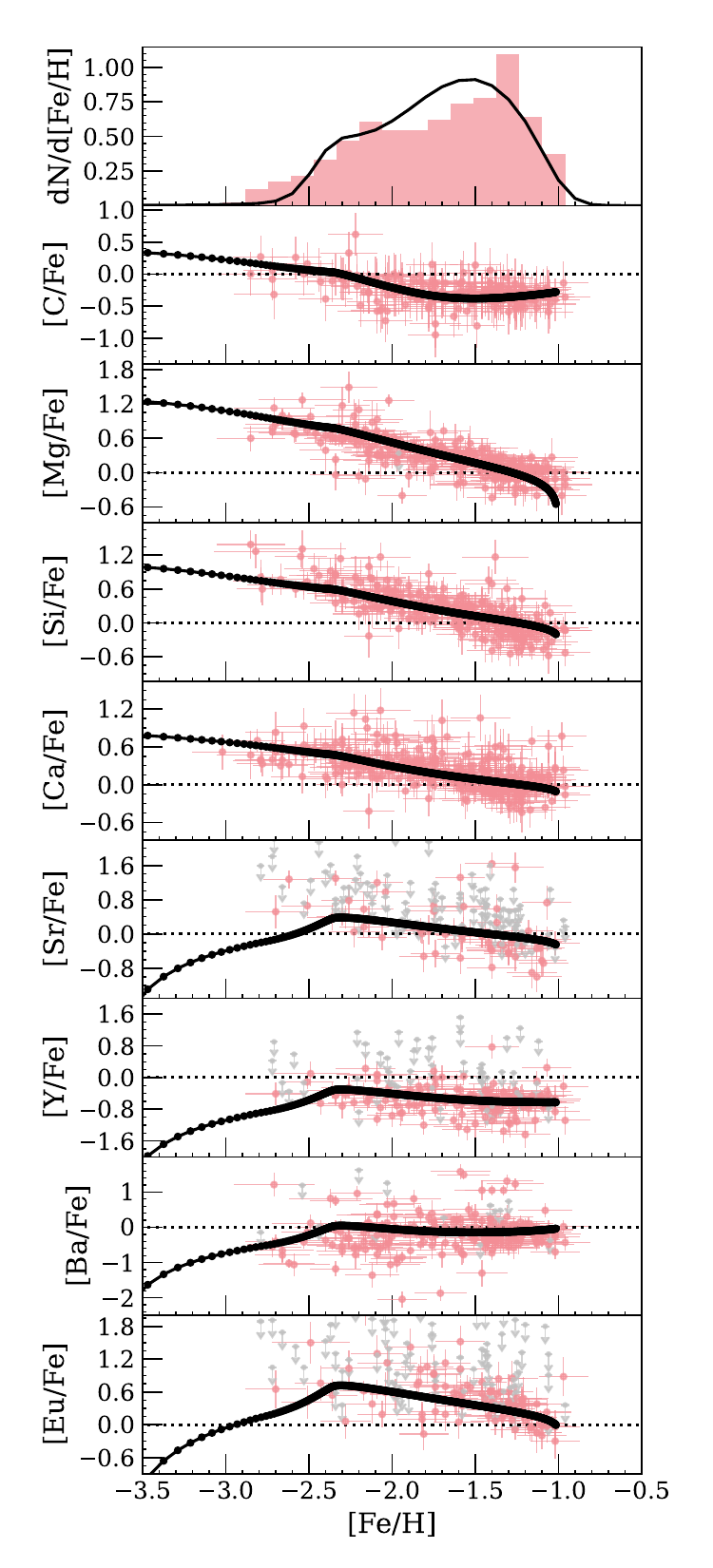}
    \caption{Metallicity distribution function (top) and [C/Fe], [Mg/Fe], [Si/Fe], [Ca/Fe], [Sr/Fe], [Y/Fe], [Ba/Fe], and [Eu/Fe] as a function of [Fe/H] for Sculptor. Pink points are abundance measurements, and gray arrows are upper limits. The black line shows the best-fit GCE model with NSMs as the only $r$-process source. This is the model that corresponds to the first column in Table \ref{tab:r_tests}.}
    \label{fig:Scl_NSMs}
\end{figure}

\begin{figure}
    \centering
    \includegraphics[width=1.0\linewidth]{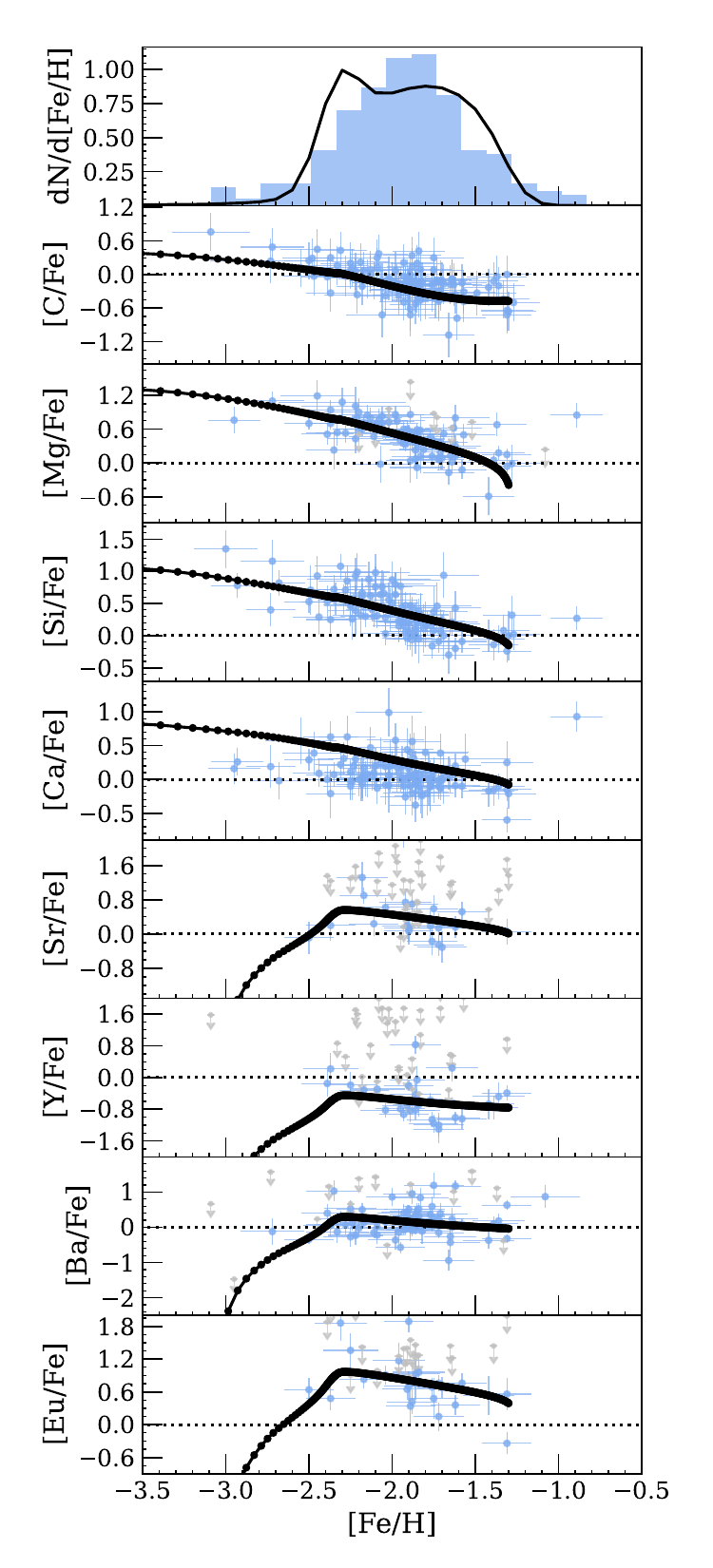}
    \caption{Same as Figure \ref{fig:Scl_NSMs}, but for the Draco dSph.}
    \label{fig:Dra_NSMs}
\end{figure}

\subsection{r-process tests} \label{subsec:rproctests}
One way to test whether NSMs are a sufficiently fast source of the $r$-process in the dSphs in our study is to allow the minimum delay time, $t_{\rm min,NSM}$, and the normalization on the NSM DTD, $A_{\rm NSM}$, to be free parameters. We performed this test for both Sculptor and Draco. In order to fit the Sr and Y abundances, we also included a normalization term for each to modify the NSM yields (more on this in Section \ref{subsec:SrandYnorms}). The free parameters fit by MCMC to the data are $A_{\rm in}$, $\tau_{\rm in}$, $A_{\rm out}$, $A_*$, $\alpha$, $t_{\rm min,NSM}$, $A_{\rm NSM}$, ${\rm normSr}_{\rm NSM}$, and ${\rm normY}_{\rm NSM}$. In these tests, and all others involving NSMs, we used the NSM yields from \citet{holmbeck_MWNSMs_2024} and the NSM DTD given in Equation \ref{eq:NSMDTD}. All the $r$-process-related tests in this section use $t_{\rm min, Ia}$ = 0.1 Gyr and $A_{\rm Ia}$ = 1.0.  The choice of $t_{\rm min, Ia}$ is motivated by the test in the previous section, and the choice of $A_{\rm Ia}$ is near the best-fit value for that test (Table~\ref{tab:Iatests}).

The results of this first test are listed in the left column of Table \ref{tab:r_tests}. The \texttt{emcee} corner plot for the fit to Sculptor can be found in Appendix \ref{appendix:corner}. The best-fitting models for  Sculptor and Draco are shown in Figures \ref{fig:Scl_NSMs} and \ref{fig:Dra_NSMs}; as these plots indicate, including NSMs in our GCE model can match the observations in both dSphs.
However, NSMs would have to occur extremely quickly after star formation: the best-fit values for $t_{\rm min,NSM}$ are 14--16 Myr. For comparison, this is the timescale for a $\sim 12~M_{\odot}$ star to evolve into a neutron star. This leaves very little time for the neutron stars to spiral together and merge, even with dynamical effects.

To further explore the timescale of the $r$-process in Sculptor and Draco dSphs, we ran two additional tests on each. In one, we set the NSM $t_{\rm min,NSM}$ to 50 Myr, an arbitrarily chosen longer delay time, and added MRSNe as an additional $r$-process source. We used MCMC to fit $A_{\rm in}$, $\tau_{\rm in}$, $A_{\rm out}$, $A_*$, $\alpha$, $A_{\rm NSM}$, $f_{\rm MRSN}$, ${\rm normSr}_{\rm MRSN}$, and ${\rm normY}_{\rm MRSN}$, with ${\rm normSr}_{\rm NSM}$ and ${\rm normY}_{\rm NSM}$ set to their best fit values from the NSM-only test. The results of this test are in the middle column of Table \ref{tab:r_tests}.

\begin{figure}
    \centering
    \includegraphics[width=1.0\linewidth]{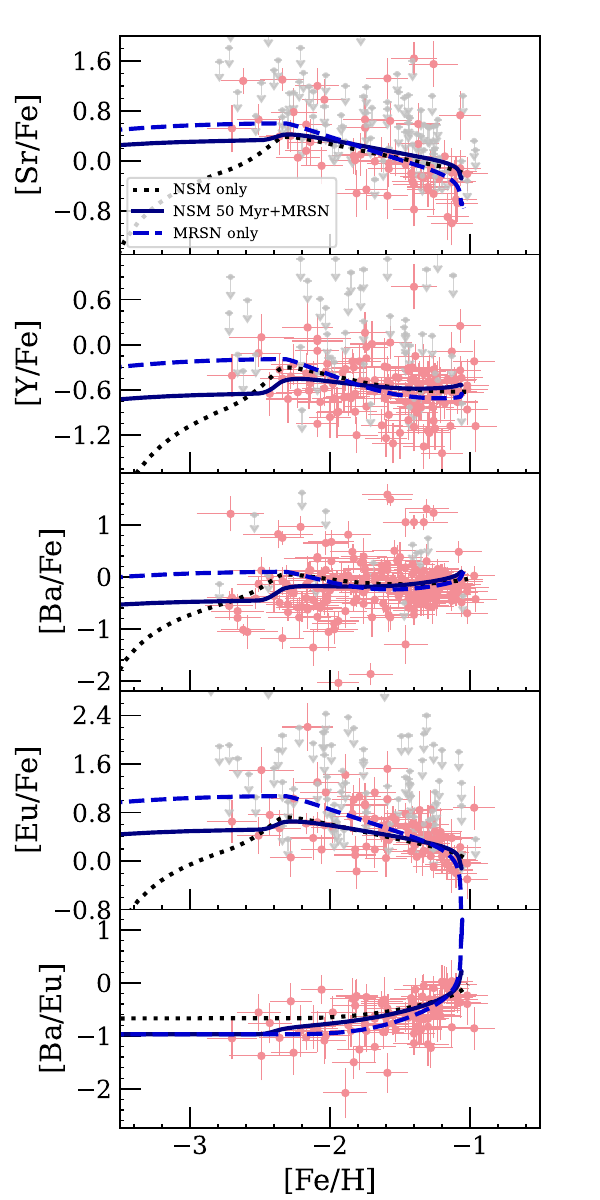}
    \caption{[Sr/Fe], [Y/Fe], [Ba/Fe], [Eu/Fe], and [Ba/Eu] as a function of [Fe/H] in Sculptor. Pink points are abundance measurements. Gray arrows are upper limits. [Ba/Eu] error bars are from [Ba/Fe] and [Eu/Fe] errors added in quadrature, with only combined errors $<$ 0.5 shown. Light gray, gray, and black lines are the best-fit GCE models for the $r$-process tests given in Table \ref{tab:r_tests}.}
    \label{fig:Scl_NSMMRSNcomparison}
\end{figure}

\begin{figure}
    \centering
    \includegraphics[width=1.0\linewidth]{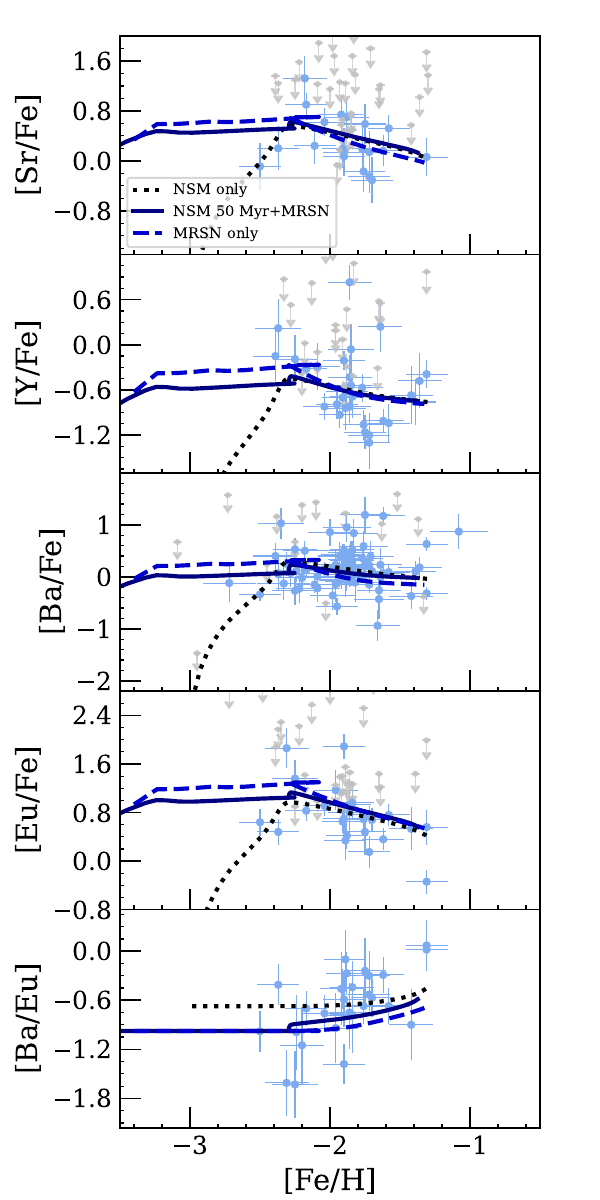}
    \caption{Same as Figure \ref{fig:Scl_NSMMRSNcomparison}, but for the Draco dSph.}
    \label{fig:Dra_NSMMRSNcomparison}
\end{figure}

In the other test, we exclude NSMs as $r$-process sources completely in order to explore whether a prompt $r$-process source associated with CCSNe could fit the $n$-capture abundance trends we observe in the galaxies. In this case, we use MCMC to fit $A_{\rm in}$, $\tau_{\rm in}$, $A_{\rm out}$, $A_*$, $\alpha$, $f_{\rm MRSN}$, ${\rm normSr}_{\rm MRSN}$, and ${\rm normY}_{\rm MRSN}$. The results of this test are in the right column of Table \ref{tab:r_tests}. All three of the best-fit models for Sculptor and Draco that are described in this section are plotted in Figures \ref{fig:Scl_NSMMRSNcomparison} and \ref{fig:Dra_NSMMRSNcomparison}.

In both tests involving MRSNe, the percentage of stars 10--35 $M_{\odot}$ that would need to explode as MRSNe, given the \citet{reichert_MRSNe_2021} yields, is less than 2\%. Clearly, including a prompt source of the $r$-process that operates on the same timescale as CCSNe can allow the NSM minimum delay time to be longer. We discuss the implications of these models on the sites of the $r$-process in detail in Section \ref{subsec:NSM_discussion}.

\begin{deluxetable}{c|c|c|c|c}
\tablecolumns{5}
\tablewidth{0pt}
\tablecaption{Best-fit parameters for Sculptor $t_{\rm min,Ia}$ tests \label{tab:Iatests}}
\tablehead{
 \colhead{Parameter} & \colhead{$t_{\rm min,Ia}$ = 50 Myr} & \colhead{$t_{\rm min,Ia}$ = 100 Myr} & \colhead{$t_{\rm min,Ia}$ = 200 Myr} & \colhead{$t_{\rm min,Ia}$ = 500 Myr}}
\startdata
$A_{\rm in}$ & $0.91^{+0.07}_{-0.12}$ & $0.76^{+0.12}_{-0.11}$ & $0.46^{+0.08}_{-0.06}$ & $0.12^{+0.02}_{-0.01}$ \\
$\tau_{\rm in}$ & $0.20^{+0.01}_{-0.01}$ & $0.22^{+0.02}_{-0.02}$ & $0.29^{+0.03}_{-0.03}$ & $0.69^{+0.09}_{-0.09}$ \\
$A_{\rm out}$ & $4.92^{+0.05}_{-0.28}$ & $4.75^{+0.18}_{-0.18}$ & $4.51^{+0.21}_{-0.20}$ & $4.07^{+0.19}_{-0.19}$ \\
$A_{*}$ & $1.28^{+0.10}_{-0.14}$ & $1.27^{+0.14}_{-0.15}$ & $0.95^{+0.11}_{-0.09}$ & $0.51^{+0.04}_{-0.04}$ \\
$\alpha$ & $0.59^{+0.02}_{-0.02}$ & $0.46^{+0.06}_{-0.05}$ & $0.46^{+0.05}_{-0.05}$ & $0.56^{+0.05}_{-0.05}$ \\
$A_{\rm Ia}$ & $0.61^{+0.04}_{-0.05}$ & $1.11^{+0.07}_{-0.07}$ & $2.48^{+0.20}_{-0.18}$ & $9.56^{+0.32}_{-0.59}$ \\
$A_{\rm NSM}$ & $3.29^{+0.17}_{-0.19}$ & $4.21^{+0.44}_{-0.44}$ & $8.68^{+0.80}_{-0.93}$ & $9.72^{+0.21}_{-0.42}$ \\
$t_{\rm min,NSM}$ (Myr) & $12^{+1}_{-1}$ & $15^{+2}_{-1}$ & $38^{+4}_{-4}$ & $64^{+4}_{-4}$ \\
${\rm normSr}_{\rm NSM}$ & $0.21^{+0.02}_{-0.02}$ & $0.19^{+0.02}_{-0.02}$ & $0.19^{+0.02}_{-0.02}$ & $0.23^{+0.02}_{-0.02}$ \\
${\rm normY}_{\rm NSM}$ & $0.03^{+0.004}_{-0.003}$ & $0.03^{+0.002}_{-0.002}$ & $0.03^{+0.002}_{-0.002}$ & $0.03^{+0.002}_{-0.002}$ \\
\hline
$-\ln{L}$ & 1521 & 1559 & 1656 & 1704 \\
\enddata
\tablecomments{Best-fit parameters quoted here are the median values from the MCMC walkers, discarding burn-in. Upper and lower uncertainty intervals are 16th and 84th percentiles from the MCMC posterior distribution. The $-\ln L$ values reflect the minimum from all the walkers in the MCMC calculation, thus representing the best fit MCMC found to the model configuration. A smaller number corresponds to a better fit to the observed data.}
\end{deluxetable}

\begin{deluxetable}{c|cc|cc|cc}
\tablecolumns{7}
\tablewidth{0pt}
\tablecaption{Best-fit parameters for $r$-process tests\label{tab:r_tests}}
\tablehead{
  \colhead{Parameter} & 
  \multicolumn{2}{c|}{NSMs only} & 
  \multicolumn{2}{c|}{NSMs with $t_{\rm min}=$50 Myr and MRSNe} & 
  \multicolumn{2}{c}{MRSNe only} \\
  \hline
  \colhead{} & 
  \colhead{Sculptor} & \colhead{Draco} & 
  \colhead{Sculptor} & \colhead{Draco} & 
  \colhead{Sculptor} & \colhead{Draco}
}
\startdata
$A_{\rm in}$ & $0.68^{+0.10}_{-0.16}$ & $1.36^{+0.21}_{-0.19}$ & $0.44^{+0.08}_{-0.05}$ & $1.33^{+0.24}_{-0.16}$ & $0.42^{+0.13}_{-0.06}$ & $1.41^{+0.20}_{-0.18}$ \\
$\tau_{\rm in}$ & $0.23^{+0.04}_{-0.02}$ & $0.16^{+0.02}_{-0.01}$ & $0.30^{+0.02}_{-0.03}$ & $0.17^{+0.02}_{-0.02}$ & $0.30^{+0.01}_{-0.05}$ & $0.16^{+0.02}_{-0.02}$ \\
$A_{\rm out}$ & $4.79^{+0.18}_{-0.18}$ & $6.76^{+0.41}_{-0.32}$ & $4.75^{+0.07}_{-0.09}$ & $6.77^{+0.36}_{-0.34}$ & $4.63^{+0.14}_{-0.18}$ & $6.61^{+0.35}_{-0.39}$ \\
$A_{*}$ & $1.18^{+0.13}_{-0.12}$ & $1.89^{+0.31}_{-0.24}$ & $1.02^{+0.10}_{-0.11}$ & $2.43^{+0.42}_{-0.41}$ & $1.11^{+0.10}_{-0.07}$ & $2.46^{+0.21}_{-0.16}$ \\
$\alpha$ & $0.49^{+0.06}_{-0.05}$ & $0.44^{+0.10}_{-0.07}$ & $0.59^{+0.04}_{-0.04}$ & $0.27^{+0.13}_{-0.09}$ & $0.52^{+0.04}_{-0.08}$ & $0.12^{+0.03}_{-0.03}$ \\
$A_{\rm NSM}$ & $3.77^{+0.30}_{-0.32}$ & $7.23^{+1.13}_{-0.89}$ & $7.26^{+0.85}_{-0.47}$ & $9.68^{+0.23}_{-0.62}$ & -- & -- \\
$t_{\rm min,NSM}$ (Myr) & $14^{+1}_{-1}$ & $16^{+2}_{-2}$ & \textbf{50} & \textbf{50} & -- & -- \\
${\rm normSr}_{\rm NSM}$ & $0.19^{+0.02}_{-0.02}$ & $0.16^{+0.03}_{-0.02}$ & \textbf{0.19} & \textbf{0.16} & -- & -- \\
${\rm normY}_{\rm NSM}$ & $0.03^{+0.002}_{-0.002}$ & $0.012^{+0.002}_{-0.001}$ & \textbf{0.03} & \textbf{0.01} & -- & -- \\
$f_{\rm MRSN}$ & -- & -- & $0.005^{+0.001}_{-0.001}$ & $0.017^{+0.002}_{-0.002}$ & $0.018^{+0.001}_{-0.002}$ & $0.03^{+0.002}_{-0.002}$ \\
${\rm normSr}_{\rm MRSN}$ & -- & -- & $0.46^{+0.13}_{-0.13}$ & $0.21^{+0.06}_{-0.05}$ & $0.24^{+0.02}_{-0.02}$ & $0.18^{+0.03}_{-0.02}$ \\
${\rm normY}_{\rm MRSN}$ & -- & -- & $0.05^{+0.01}_{-0.01}$ & $0.02^{+0.005}_{-0.004}$ & $0.04^{+0.004}_{-0.003}$ & $0.02^{+0.002}_{-0.002}$ \\
\hline
$-\ln{L}$ & 1562 & 940 & 1527 & 951 & 1545 & 961 \\
\enddata
\tablecomments{Values in this table follow the same format as Table \ref{tab:Iatests}. Bold values were not free parameters in the model fit.}
\end{deluxetable}

\subsection{Error quantification} \label{subsec:error}
\begin{figure}
    \centering
    \includegraphics[width=1.0\linewidth]{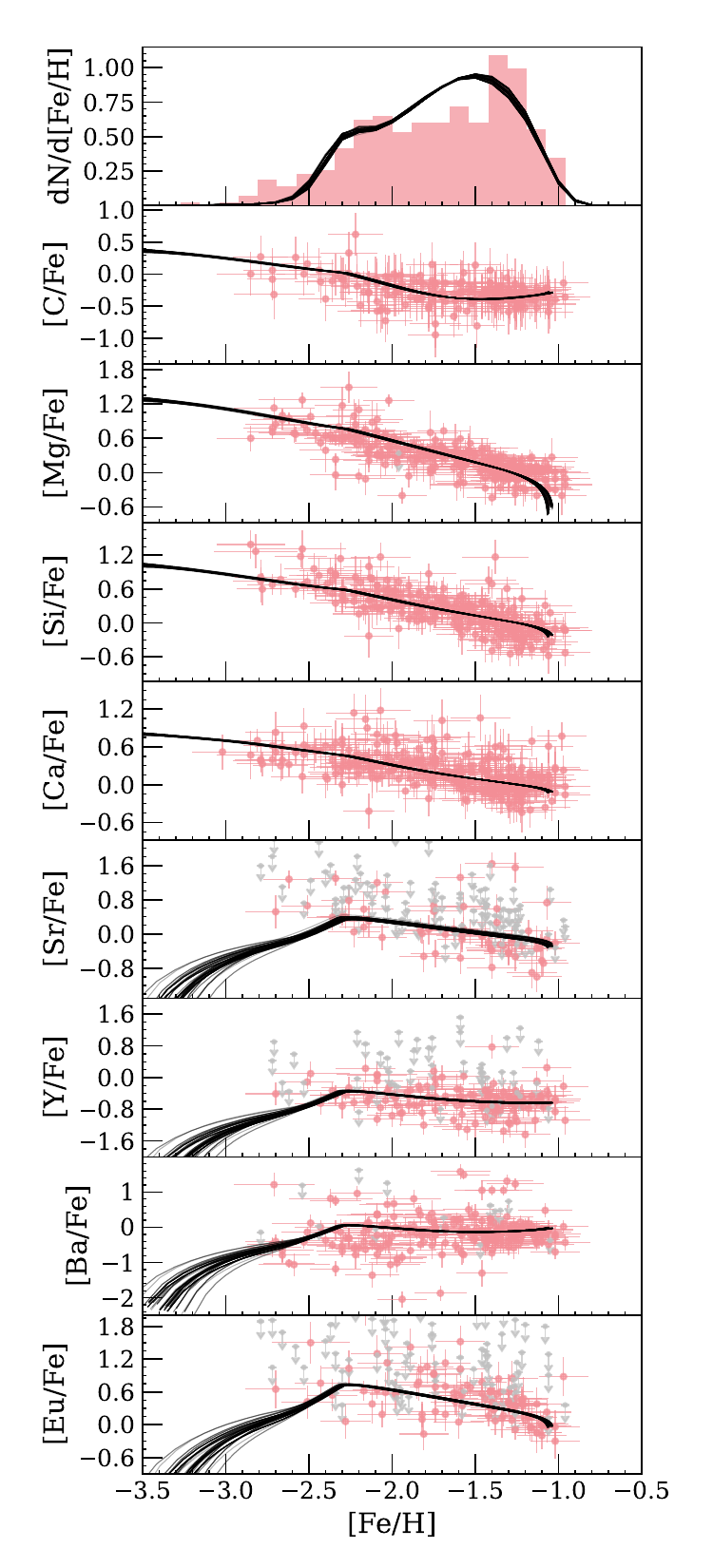}
    \caption{Metallicity distribution function (top) and [C/Fe], [Mg/Fe], [Si/Fe], [Ca/Fe], [Sr/Fe], [Y/Fe], [Ba/Fe], and [Eu/Fe] as a function of [Fe/H] for Sculptor. Pink points are abundance measurements, and gray arrows are upper limits. The black lines are 100 model realizations randomly chosen from the MCMC, which fits $A_{\rm in}$, $\tau_{\rm in}$, $A_{\rm out}$, $A_{*}$, $\alpha$, $A_{\rm NSM}$, $t_{\rm min,NSM}$, ${\rm normSr}_{\rm NSM}$, and ${\rm normY}_{\rm NSM}$, excluding the 5000 burn-in steps. The best-fit parameters from the MCMC are given in the first column of Table \ref{tab:r_tests}.}
    \label{fig:error}
\end{figure}

The upper and lower errors quoted in Tables \ref{tab:Iatests} and \ref{tab:r_tests} are 16th and 84th percentiles from the MCMC posterior distribution, excluding burn-in steps. Figure \ref{fig:error} shows 100 random walker steps from our MCMC run for Sculptor in which NSMs are the only $r$-process sources (corresponding to the first column in Table \ref{tab:r_tests}). This plot gives a very basic idea of the parameter space the walkers explored. All 100 models fit the [$\alpha$/Fe] measurements well while deviating in their $n$-capture elements abundances at low metallicities. This shows that the $n$-capture abundances are adding new constraints on the model fits beyond [$\alpha$/Fe]. We discuss this degeneracy-breaking further in Section \ref{subsec:Ia_discussion}.

Our procedure does not provide the full picture of model uncertainties. As with any GCE model, the parameterization of DTDs, star formation, and gas inflow and outflow represent significant sources of error. The assumed IMF also effects the GCE models, as it determines the relative numbers of high and low mass stars. This can change [X/Fe] evolution in the model as well as the inferred SFH \citep{delosreyes_simultaneous_2022, palla_SMC_2026}. For a detailed look into GCE model dependence on the IMF, SNIa DTD, and gas/stellar mass, see \citet{cote_uncertainties_2016}. There are also sources of error beyond the formulation of the GCE model that come from stellar models and nucleosynthesis yields \citep{romano_quantifying_2010, jones_massive_star_evolution_2015, molla_galactic_2015, mumpower_nuclearmasses_2015, kobayashi_SNIayields_2020}. These sources of error are difficult to estimate because of the complex interdependence of the model variables. Thus, we find quoting the MCMC percentile errors to be the simplest solution. 

\section{Discussion} \label{sec:discussion}

\subsection{Fits to the MDFs} \label{subsec:MDFs}
Our GCE models fit some aspects of the observed [X/Fe] vs.\ [Fe/H] trends in Sculptor and Draco (Figures \ref{fig:Scl_NSMs} and \ref{fig:Dra_NSMs}) well. The models reproduce the approximate MDF and the trends after the [$\alpha$/Fe] v. [Fe/H] ``knee.''  The reasonably good fits in this space indicate that we have captured the qualitative ratio of SNIa rates to the rates of more prompt sources of elements.

Draco's MDF is not particularly well-fit by any of our GCE models, regardless of the SNIa or $r$-process prescriptions we use. In particular, Draco's MDF is more sharply-peaked than what the model can accomplish while simultaneously fitting the stellar [X/Fe] abundances. This is likely because the parameterizations for gas inflow and outflow (Equations \ref{eq:inflow} and \ref{eq:outflows}) that work well for Sculptor are not applicable to Draco.  Even Sculptor has a similar incongruity in its narrow [Fe/H] peak at ${\rm [Fe/H]} = -1.3$. More detailed investigation of inflow and outflow parameterization will be included in a future paper.

Interestingly, \citet{cote_GCEassumptions_2017} found that the three parameterizations of inflow and outflow that they tested did not make a significant difference in their MCMC fits to the [X/Fe] abundances in Sculptor. Their analysis did not include any neutron-capture abundances in Sculptor, so perhaps the $s$-process can provide a useful constraint to the choice of evolutionary model for dSphs.

\subsection{SNIa DTD} \label{subsec:Ia_discussion}
Many of the studies of the SNIa DTD come from direct observations of supernovae in distant galaxies \citep[e.g.,][]{gal-yam_ia_rate_2002, maoz_IaDTD_2010, barbary_ia_rate_2012}.  The measurement of the DTD is indirect because each galaxy has at most one observed supernova.  No single galaxy can give a complete DTD\@.  The DTD must be inferred statistically from the observed galaxy population-level rates and the derived SFHs of the host galaxies.

Alternatively, the DTD can be measured from GCE models \citep[e.g.,][]{kobayashi_ia_rate_1998}.  A GCE model can relate the metallicity of stars in a galaxy to the time at which those stars were born, thus establishing a delay time for the production of iron from SNeIa.  The DTD can be varied until the GCE model matches the data.

Previous GCE studies of dSphs have shown that the minimum delay time of SNeIa and the SFH of the galaxy are degenerate when only iron-group and $\alpha$ elements are included in the model (see Figure 10 of \citealt{kirby_multi-element_alpha_2011}). When the onset of SNeIa is faster, a shorter SFH compensates and can reproduce the same MDF and $\alpha$-Fe track as if the SNeIa minimum delay were longer. This is illustrated by the [Mg/Fe] vs.\ [Fe/H] panel in Figure \ref{fig:Scl_Iatmin}. All four models fit [Mg/Fe] equally well by requiring different SFHs. Most studies address this by fixing either the SNIa DTD \citep{weinberg_equilibrium_2017, delosreyes_simultaneous_2022} or fixing the SFH \citep{deboer_sculptor_2012, heiger_dleiy_2026}. Another way to break this degeneracy is to include nucleosynthesis from another source, such as the $s$-process.  The new source serves as an independent chemical clock, relieving SNeIa from solely determining the age--metallicity relation.

The $s$-process nucleosynthesis from AGB stars acts as another lever to constrain the timescale in GCE models because the time it takes AGB stars to evolve, and subsequently enrich the ISM with $s$-process material through their winds, is well-known \citep{gallino_AGBs_1998, lugaro_s-process_2003, karakas_slar_2016, karakas_SMCAGBs_2018} \footnote{This is not to say that there are no open questions regarding $s$-process nucleosynthesis. See \citet{lugaro_s_and_beyond_2023} for detailed $s$-process discussions.}. Though the SFHs of Sculptor and Draco are too short to include enrichment from lower mass AGBs (1-2 $M_{\odot}$), the $s$-process in higher-mass AGBs is still an important contributor in the models. By including Sr, Y, and Ba in our models we are able to break the degeneracy between the SNIa minimum delay time and the SFH, thus placing stronger constraints on the GCE timescale in the dSphs. 

\begin{figure}
    \centering
    \includegraphics[width=1.0\linewidth]{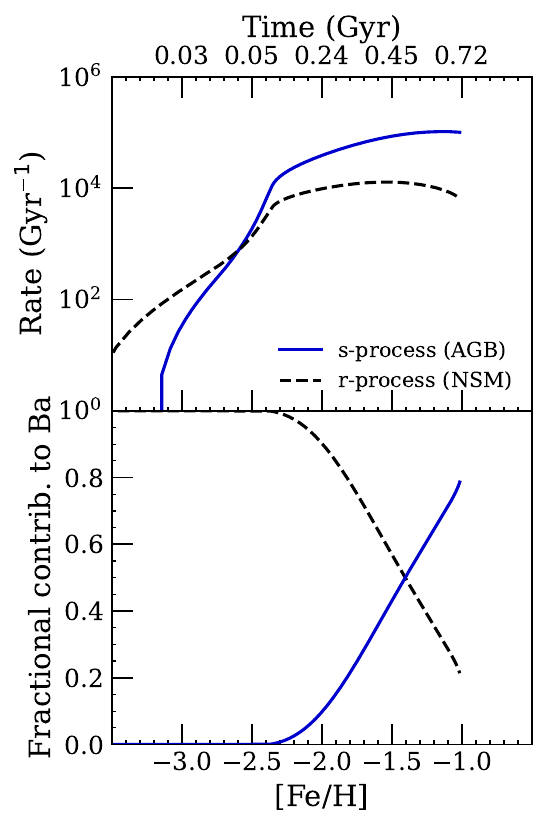}
    \caption{Top: $s$-process (blue, solid line) and $r$-process (black, dashed line) event rates in Sculptor as a function of [Fe/H] on the lower x-axis and time from the start of the model on the upper x-axis. Bottom: Fractional contribution to the Ba enrichment at each timestep from $s$-process and $r$-process as a function of [Fe/H] and time in Sculptor. The model used in both panels of this figure is given in the first column of Table \ref{tab:r_tests}, with only NSMs as $r$-process sites. }
    \label{fig:AGBcontrib}
\end{figure}

These elements are also produced by the $r$-process in different events with completely different DTDs, so it is important to have an estimate of the $s$/$r$ ratio for each element.  This estimate can be provided by ratios of Eu, made almost entirely in the $r$-process, to the other neutron-capture elements.  Our GCE model includes both $r$- and $s$-process sources and predictions for Eu abundances.  Therefore, it ``knows'' how much of each element to assign to $s$-process sources. Figure \ref{fig:AGBcontrib} shows $s$- and $r$-process event rates and their fractional contribution to Ba as a function of metallicity and time in Sculptor. The model depicted is the fit to Sculptor with only NSMs contributing to the $r$-process (the first column in Table \ref{tab:r_tests}), thus AGBs are fully responsible for the $s$-process rate and NSMs for the $r$-process. This figure illustrates how Eu abundances allow the $s$- and $r$-process contributions to Sr, Y, and Ba to be disentangled.

Figure \ref{fig:Scl_Iatmin} shows our GCE model with 4 different SNIa $t_{\rm min}$ values, ranging from 0.05 to 0.5 Gyr. The details of these tests can be found in Section \ref{subsec:Iatests}. With $s$-process elements included, the best-fit models for each of the $t_{\rm min,Ia}$ values do not all fit the MDF of Sculptor equally well. This means that the stellar abundances, when taken together, contain enough information to constrain the SFH and $t_{\rm min,Ia}$ simultaneously. 

These tests all assume the SNIa DTD power-law slope to be $-1.1$ (Equation \ref{eq:IaDTD}, \citealt{maoz_IaDTD_2010}). There are multiple reasons we chose not to formally test different SNIa DTD power-law slopes in this paper. The first is that there is some degeneracy between the slope of the SNIa DTD, the minimum delay time, $t_{\rm min,Ia}$, and the overall shape of the SNIa DTD (e.g., exponential, power-law, slow start) \citep{palicio_SNIa_MW_2023, dubay_extended_2024}. If the SNIa DTD is parameterized with a flat start followed by a power-law (as in \citealt{rajamuthukumar_triple_system_2023}), it has the same effect as a power-law DTD with a different normalization and $t_{\rm min,Ia}$. When we ran an MCMC test with $t_{\rm min,Ia}$, $A_{\rm Ia}$, and the power-law slope as free parameters we found that the data are not able to constrain all three at once. Thus, it was necessary to choose one to fix; we chose the slope. 

The other reason we chose to fix the SNIa DTD power-law slope for all of the tests is that the dSphs Sculptor and Draco are not very sensitive to the shape of the SNIa DTD\@. Under reasonable assumptions of the DTD parameters,  Sculptor and Draco's derived chemical SFH lasts less than 1~Gyr. Different SNIa DTD shapes are best distinguished by their behavior at late times, something that is much better suited to tests with Milky Way stars \citep[e.g.,][]{dubay_extended_2024}. When $t_{\rm min,Ia}$ is a free parameter, essentially any SNIa DTD shape can be made to fit by changing the rate of SNIa. Our GCE model is sensitive to the overall SNIa timescale, but not to the details of the SNIa DTD\@.

If we take the SNIa DTD to be described by Equation \ref{eq:IaDTD}, we can infer $t_{\rm min,Ia}$ from our tests shown in Figure \ref{fig:Scl_Iatmin}. The $\ln L$ values in Table \ref{tab:Iatests} suggest that $t_{\rm min,Ia}$ = 0.05 Gyr and 0.1 Gyr fit the data the best. We chose to use $t_{\rm min,Ia}$ = 0.1 Gyr for the rest of the models in this paper to match dlR22. The best-fit parameters for the $t_{\rm min,Ia}$ = 0.05 and 0.1 Gyr models are consistent within 1 $\sigma$ (except for $A_{\rm Ia}$).

The minimum delay time of SNeIa should roughly correspond to the lifetime of the most massive stars that will explode as SNeIa.  The most massive stars that become white dwarfs have masses of about $8~M_{\odot}$, corresponding to a lifetime of about 30~Myr \citep{padovani_massloss_1993,kodama_stellar_lifetime_1997}.   A SNIa minimum delay time of 0.1 Gyr---an inference that would be different with a different DTD power-law slope---corresponds to the time it takes a $4~M_{\odot}$ star to become a white dwarf.  If no SNeIa explode before 0.1~Gyr, then there are two possible implications for the progenitor stars.  First, in the single-degenerate scenario where a secondary star transfers mass to a primary white dwarf \citep{whelan_snia_1973}, the secondary (donor) star cannot be more massive than $4~M_{\sun}$.  If it were, then the minimum delay time would be shorter.  Second, in the double-degenerate scenario \citep{iben_dd_1984} where two white dwarfs merge, the sum of the time to form the white dwarfs and their inspiral times must be at least 0.1~Gyr.  In the unrealistic limit of negligible inspiral time, the minimum delay time would imply that the secondary white dwarf must be less massive than $4~M_{\odot}$.  In the opposite limit, the minimum amount of time for a white dwarf to form is 30~Myr, which means that the minimum inspiral time must be at least 70~Myr (the difference between 0.1 Gyr and 30~Myr).  That mandatory inspiral time would rule out near-equal-mass ultra-compact white dwarf binaries as an immediate product of common envelope evolution.  The merger time in such binaries are much shorter than the time it takes to evolve to a white dwarf, possibly even less than 1~Myr \citep{Mennekens_IaDTD_2010}.

Our inferred minimum delay time is shorter than the 0.5~Gyr estimated by \citet{homma_chemical_2015}.  They used only [$\alpha$/Fe] without neutron-capture elements and the SFH measured by \citet{deboer_sculptor_2012}.  That derivation of SFH is longer than most photometric studies of Sculptor's SFH \citep{weisz_star_2014,bettinelli_sculptor_2019}.  A minimum SNIa delay time of 0.5~Gyr is in strong tension with studies of galaxy clusters, albeit measured at different metallicities \citep[see][]{kobayashi_ia_rate_1998}.  Our inclusion of neutron-capture elements results in a shorter SFH, closer in agreement with the photometric studies.  As a result, the minimum delay time is less in tension with studies of SNIa rates.
For example, \citet{chen_IaDTD_2021} estimated $t_{\rm min,Ia}$ = 0.12 Gyr from observations of spatially-resolved SNIa host galaxies.

For this paper, placing actual constraints on $t_{\rm min,Ia}$, and thus on the SFH, is important because it allows us to make real estimates of the $r$-process timescale from our model. If $t_{\rm min,Ia}$ and the SFH were completely unconstrained, the minimum delay time for various $r$-process events would not be constrained either (see the $t_{\rm min,NSM}$ row in Table \ref{tab:Iatests}). In the next sections, we will discuss these results for the $r$-process.

\subsection{NSM minimum delay time} \label{subsec:NSM_discussion}
The MCMC tests described in Section \ref{subsec:rproctests} place useful constraints on the $r$-process minimum delay time in dSphs. The best fit models with $r$-process enrichment solely from NSMs for Sculptor and Draco are shown in Figures \ref{fig:Scl_NSMs} and \ref{fig:Dra_NSMs}. [Eu/Fe] at low metallicities (${\rm [Fe/H]} < -2$) places the strongest constraint on the $r$-process delay time, but Sr, Y, and Ba at low metallicity help as well. At the galaxy age corresponding to ${\rm [Fe/H]} < -2$, AGB stars have not had time to contribute significant amounts of the $s$-process, so all the neutron-capture elements are almost entirely being produced in the $r$-process (Figure \ref{fig:AGBcontrib}).

If NSMs are the only source of the $r$-process in Sculptor and Draco, they must begin after only 14--16 Myr. This corresponds to the time it takes a $12~M_{\odot}$ star to evolve into a neutron star. In order for NSMs alone to account for the $r$-process abundances in Sculptor and Draco, two neutron stars must form, inspiral, and the resulting material must mix into the ISM in just 14--16~Myr. This is not physically realistic unless common envelope evolution creates neutron stars already on the cusp of merging.

In a recent study, \citet{chattaraj_nsmdelaytime_2026} modeled binary neutron star populations over a range of metallicites and found that the vast majority of NSMs occur 40 Myr after formation of the binary neutron star system. Very few outliers occurred within $~20$~Myr of formation. Likewise, \citet{zevin_sGRB_NSMdelaytime_2022} estimated a NSM minimum delay time of 184 Myr based on observations of short gamma-ray bursts that are likely associated with NSMs.

Our finding of a short $r$-process delay time also corroborates recent studies of the Milky Way's chemical evolution. \citet{molero_MWrprocess_2026} found that GCE models of the Milky Way fit the stellar $r$-process abundances best with a short $r$-process delay time of $\leq$ 30 Myr. There are also a plethora of low-metallicity stars in the Milky Way's halo that are highly $r$-process enhanced \citep[e.g.,][]{cain_rIII_2020, placco_peculiarstar_2020, bandyopadhyay_rII_2024, shah_rstar_2024}. Clearly, something is producing the $r$-process shortly after the onset of star formation.


\subsection{NSM yields} \label{subsec:NSMyields_discussion}
The GCE tests for Sculptor and Draco also constrain the average $r$-process yields per NSM, assuming NSMs are contributing all of the $r$-process in the galaxies. Estimates of $r$-process ejecta from NSMs in the literature range from $\mathcal{O}(-4)$ - $\mathcal{O}(-2)$ $M_{\odot}$ per event, depending on the neutron star equation of state, neutron star masses, binary eccentricity, and merger physics \citep{bauswein_NSMejecta_2013, hotokezaka_NSMejecta_2013, sekiguchi_NSMejecta_2015, radice_NSMmassejection_2016, papenfort_eccentricNSMs_2018, radice_NSMejecta_2018}. Thus, it is useful to place some observational constraints on the amount of $r$-process material produced per event in dSphs.

In this work, we used the yields of \citet{holmbeck_MWNSMs_2024} as a baseline, with Sr and Y values scaled (see Section \ref{subsec:SrandYnorms}). They estimated that the average MW binary neutron star system will produce about $10^{-7}~M_{\odot}$ of Eu when they merge. These yields would require $\sim7000$ ($\sim8000$) NSMs over the course of Sculptor's (Draco's) star formation to fit the observed [Eu/Fe] abundances based on the model parameters given in column 1 (2) of Table \ref{tab:r_tests}. Considering the same model only produced $\sim4000$ CCSNe over the same period of time, this is not a realistic number of NSMs. However, scaling the normalization of the NSM DTD (Equation \ref{eq:NSMDTD}), $A_{\rm NSM}$, is the same as scaling the yields per NSM\@. If the yields are increased, fewer NSMs are needed.

If instead we require only $\sim1\%$ of CCSNe to eventually result in a NSM, only $\sim40$ NSMs would occur while Sculptor was forming stars. This corresponds to an average yield of $\mathcal{O}(-5)$ $M_{\odot}$ Eu per event, which does match some estimates in the literature \citep{wanajo_NSMyields_2014, just_NSMnucleosynthesis_2015, radice_NSMmassejection_2016, papenfort_eccentricNSMs_2018}.

\subsection{A prompt $r$-process source} \label{subsec:promptr}
We chose to use MRSNe as the representative prompt $r$-process source in this study. Any $r$-process source that occurs concurrently with star formation could take the place of MRSNe, given that it produces the $r$-process up to the lanthanides (including Eu) and in sufficient amounts. We also do not want to exclude the possibility that there are multiple prompt $r$-process channels. The abundances in stars in dSphs are not detailed enough to distinguish between $r$-process sources with similar DTDs, but they are very good at constraining the timescale of $r$-process enrichment.

In Section \ref{subsec:rproctests} we describe our MCMC tests that include both NSMs and MRSNe. We set the NSM minimum delay time to a realistic, arbitrary value of 50 Myr \citep{chattaraj_nsmdelaytime_2026} and test whether MRSNe can make up the lack of $r$-process enrichment from NSMs at early times. The results of these tests appear in the middle column of Table \ref{tab:r_tests}. The $\ln L$ values indicate that this scenario fits the Sculptor abundances better than the NSM-only scenario, but has the opposite effect for Draco.

The fraction of 10--35~$M_{\odot}$ stars that must produce $r$-process elements in Sculptor and Draco in these tests are $0.5\%$ and $1.7\%$, respectively. This corresponds to 19 MRSNe in Sculptor and 48 MRSNe in Draco over the course of their evolution. For reference, in their GCE models of the Milky Way, \citet{molero_MWncapture_2023} set $20\%$ of 10--25 $M_{\odot}$ stars to explode as MRSNe using MRSN yields from \citet{nishimura_MRSN_2017}. \citet{kobayashi_origin_2020} assumed $1.5\%$ of 25--50 $M_{\odot}$ stars explode as MRSNe with yields from \citet{nishimura_r-process_2015}. As with NSMs (Section \ref{subsec:NSMyields_discussion}), the $r$-process yields from MRSNe and the percentage of massive stars that explode as MRSNe are degenerate. Larger yields per event require fewer events. \citet{woosley_GRB_2006} estimated that only $\sim1\%$ of massive stars would have the right properties at the end of their lives to explode as MRSNe.

From these tests, we conclude that if NSMs are more delayed than the $t_{\rm min,NSM}$ = 14 Myr required of NSMs in our models,  a prompt $r$-process source with the same timescale as massive star evolution is able to explain the $r$-process observations at early times. This disagrees with the results of \citet{duggan_neutron_2018}, who concluded that the majority of the $r$-process in dSphs is from NSMs. The difference is likely because they did not have direct Eu abundances for their stars, while we have the benefit of a large sample of Eu abundances. On the other hand, our results corroborate those of \citet{skuladottir_evidence_2020}, who found that the Milky Way and dSph $r$-process abundances fit best with both a fast $r$-process component on the same timescale as CCSNe and a significantly delayed component, which they attribute to NSMs.

We also ran MCMC tests to see whether MRSNe alone could reproduce the $r$-process enrichment in Sculptor and Draco. If the only $r$-process source is MRSNe, then $1.8\%$ of 10--35~$M_{\odot}$ stars in Sculptor and $3\%$ in Draco must explode as MRSNe. It is not surprising that MRSNe alone can explain the $r$-process abundances in the dSphs almost as well as (unrealistically prompt) NSMs alone. In the models with only NSMs producing the $r$-process, the NSM minimum delay time is so short that the onset of NSMs is essentially the same as the onset of MRSNe. The difference between the DTDs of NSMs and MRSNe in these models is that NSMs have a longer tail of delayed events (Equation \ref{eq:NSMDTD}), whereas MRSNe follow the CCSN timescale and are thus governed by stellar evolution and the IMF\@.

Comparisons of the best-fit GCE models with and without MRSNe are shown in Figures \ref{fig:Scl_NSMMRSNcomparison} and \ref{fig:Dra_NSMMRSNcomparison}. For Sculptor, the models that include MRSNe more easily account for the high [Sr/Fe], [Y/Fe], and [Eu/Fe] at [Fe/H] $\leq$ -2.5.

Our tests indicate that the first $r$-process source was prompt, and most likely more prompt than NSMs.  Our models still allow for copious delayed NSMs as long they are not the only source of the $r$-process.  After all, the only directly confirmed $r$-process site is NSMs.


\subsection{Sr and Y yield overestimation} \label{subsec:SrandYnorms}
\begin{figure*}
    \centering
    \includegraphics[width=1.0\linewidth]{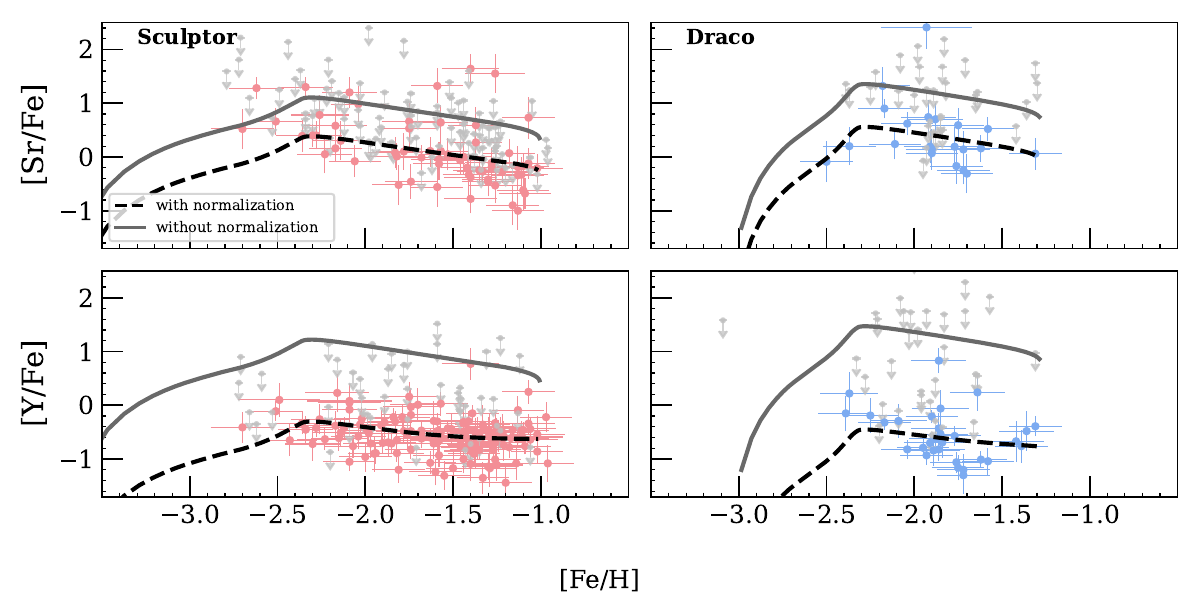}
    \caption{[Sr/Fe] and [Y/Fe] as a function of [Fe/H] for Sculptor (left, pink) and Draco (right, blue). Pink and blue points are abundance measurements from DEIMOS, and gray arrows are upper limits. The black line is the best-fit GCE model with NSMs supplying the $r$-process (the same model as in Figures \ref{fig:Scl_NSMs} and \ref{fig:Dra_NSMs}). The gray lines are GCE models with all of the same parameters, but ${\rm normSr}_{\rm NSM}$ and ${\rm normY}_{\rm NSM}$ are set to 1.}
    \label{fig:Scl_SrYnorm}
\end{figure*}

The theoretical $r$-process yields from NSMs and MRSNe we used in this paper \citep{holmbeck_MWNSMs_2024, reichert_MRSNe_2021} report too much light $r$-process contribution (Sr and Y) compared to Ba and Eu to fit the observed abundances.
Figure~\ref{fig:Scl_SrYnorm} shows the prediction of our model without scaling down the Sr and Y yields from NSMs, compared to models with the rescaling.  Our models require Sr and Y from NSMs to be reduced by a factor of $\sim$ 5 and $\sim$ 50, respectively (values of ${\rm normSr_{NSM}}$ and ${\rm normY_{NSM}}$ are in Table \ref{tab:r_tests}). These yield scaling values should be taken as qualitative, as their exact values depend on model assumptions such as the IMF.

While the need for Sr and Y yield scaling factors adds additional terms we must account for when fitting the models, it is not necessarily surprising. In NSMs the final yields of light $r$-process elements are highly dependent on the physics in the disk after the neutron stars merge \citep{fernandez_2019_nsmgrmhd,qiu_nsmnu_2025,lund_2025_nsmnu,froustey_2026_nsmnu,bernuzzi_nsmhydro_2026}. For example, \citet{sprouse_NSMdisk_2024} found a reduction of over a half of an order of magnitude in the light $r$-process element yields relative to the main $r$-process yields of their NSM models when the accretion disk surrounding the remnant black hole was modeled for longer. 

The reduction in the MRSNe Sr and Y yields relative to Ba and Eu that our models require is more difficult to reconcile with astrophysical simulations. MRSNe models in general show significantly more weak $r$-process production than main, with robust Eu yields found only in the models with the strongest magnetic fields \citep{nishimura_2017_mhdsne,mosta_2018_mhdsne,reichert_MRSNe_2021}. Inferring MRSNe to be the prompt $r$-process source in Sculptor and Draco is therefore in tension with our results. Such tension is likely also present with other possible prompt sources such as collapsars and magnetar giant flares that are similarly expected to produce more weak $r$-process than main.

A possible prompt source that may robustly produce lanthanides with dramatically less Sr and Y than an $r$-process is the recently-explored $\nu i$-process \citep{wang_2026_nui}. In a $\nu i$-process, lanthanides are produced via late-time neutron captures in \emph{proton}-rich conditions dominated by high neutrino fluxes, as can be found in rare, unusually-energetic core collapse events (hypernovae). While the $\nu i$-process is still speculative, and its potential yields depend sensitively on uncertain neutrino physics \citep{balantekin_2025_nui}, it may provide a ratio of Sr and Y to Ba and Eu more in line with our models.

\section{Conclusion} \label{sec:conc}
The main conclusions of this paper are as follows:
\begin{enumerate}
    \item If NSMs are the only $r$-process source in dSphs, they must begin just 14--16 Myr after the onset of star formation.
    \item Prompt $r$-process sources that operate on the same timescales as CCSNe are able to reproduce the neutron-capture element abundances in Sculptor and Draco, either on their own or paired with NSMs. Our models require just 0.5 -- 3\% of 10--35 $M_{\odot}$ stars to produce the  $r$-process by dying as MRSNe.
    \item Including both $s$- and $r$-process enrichment in the GCE model breaks the degeneracy between the SNIa minimum delay time and the SFH of the galaxy. This makes it possible to draw stronger conclusions about the timescales of metal enrichment in GCE models.
    \item Our models with NSMs require modifications to the yields of \citet{holmbeck_MWNSMs_2024} that we used. First, the yields of Ba and Eu per NSM event must increase by a factor of 10--100 to allow for a realistic number of NSMs to occur in Sculptor and Draco. Second, the Sr and Y yields must be decreased in relation to Ba and Eu by a large amount to avoid overproducing Sr and Y compared to the observed abundances.
\end{enumerate}

Our comparisons between GCE models and neutron-capture stellar abundances in Sculptor and Draco are made possible by the large metallicity range and sample size attained with medium-resolution multi-object spectroscopy. This work, while focused on two of the most massive dSphs in the Local Group, sets the stage for complementary work on smaller dSphs. Instruments such as the Subaru '$\bar{\rm O}$nohi'ula Prime Focus Spectrograph \citep{takada_pfs_2014, chiba_PFSGAsurvey_2026}, 4MOST \citep{dejong_4most_2019}, and VLT/MOONS \citep{cirasuolo_moons_2020} will make it possible to study the chemical evolution of Local Group galaxies in even greater detail. In particular, comparisons between these results and other dSphs like Sextans and Ursa Minor should be done to test whether the timescale of the $r$-process is consistent among dSphs.

\begin{acknowledgments}
Some of the data presented herein were obtained at Keck Observatory, which is a private 501(c)3 non-profit organization operated as a scientific partnership among the California Institute of Technology, the University of California, and the National Aeronautics and Space Administration. The Observatory was made possible by the generous financial support of the W. M. Keck Foundation.  The authors wish to recognize and acknowledge the very significant cultural role and reverence that the summit of Maunakea has always had within the Native Hawaiian community. We are most fortunate to have the opportunity to conduct observations from this mountain.

R. S. acknowledges support from the U.S. Department of Energy under Grant Nos. DE-FG02-95-ER40934 and DE-SC0024388 (ENAF) and the National Science Foundation under Grant Nos. PHY-2020275 (N3AS) and 21-16686 (NP3M).

\end{acknowledgments}

\begin{contribution}

LEH edited the GCE model, calculated the models, and wrote the manuscript. ENK and MdlR developed the GCE model and edited the manuscript. RS edited the manuscript and contributed to interpretation of the results.


\end{contribution}

%
\facility{Keck:II (DEIMOS)}

\software{\texttt{emcee} \citep{emcee_2013}
\texttt{matplotlib} \citep{Hunter_matplotlib_2007}
          }

\begin{appendix}
\section{Fornax} \label{appendix:fornax}
The Fornax dSph galaxy is the most massive of the Local Group dSphs. It also has a prolonged, complex SFH \citep{weisz_star_2014, rusakov_bursty_2021} and a large spread in abundances at a given metallicity, indicative of an ISM that is not well-mixed as stars are forming. For these reasons, the simple GCE model presented in this paper is not well-suited to reproducing the abundances in Fornax, as opposed to the much more ``well-behaved'' dSphs Draco and Sculptor.

Even so, we applied our model with neutron-capture nucleosynthesis to the abundance measurements in Fornax. The abundances are all from the same sources listed in Section \ref{subsec:abundances}. In Table \ref{tab:fornax} we provide the best-fit model parameters for a fit to the Fornax abundances where NSMs are the only $r$-process source (analogous to the models for Sculptor and Draco in the first column of Table \ref{tab:r_tests}). Figure \ref{fig:Fornax_model} shows the GCE model plotted over the Fornax abundances. Due to the assumption of instantaneous mixing in our model, the spread in the $\alpha$  elements is not reproduced. 

\begin{figure}
    \centering
    \includegraphics[width=0.75\linewidth]{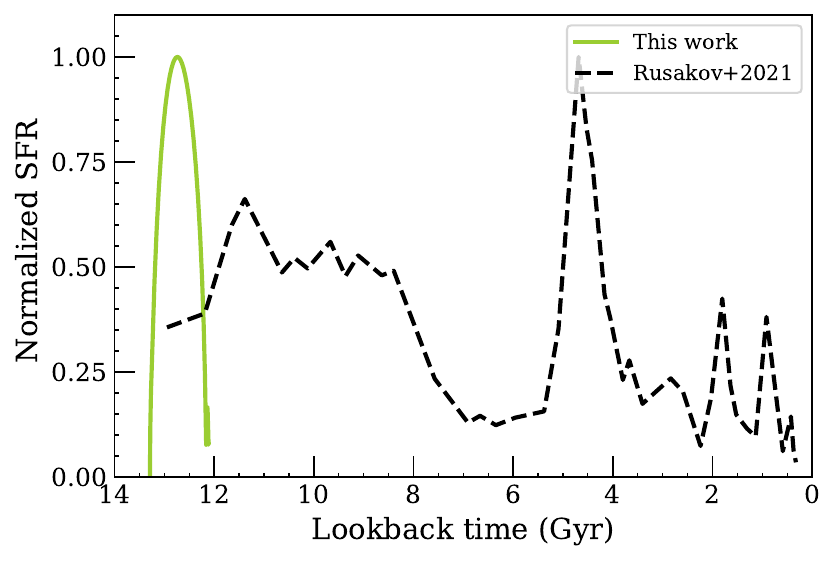}
    \caption{Star formation rate normalized to its maximum value as a function of lookback time for the Fornax dSph. The green line is the SFH from our GCE model (parameters given in Table \ref{tab:fornax}), while the dashed black line is a SFH from \citet{rusakov_bursty_2021}.}
    \label{fig:FornaxSFH}
\end{figure}

\begin{figure}
    \centering
    \includegraphics[width=0.5\linewidth]{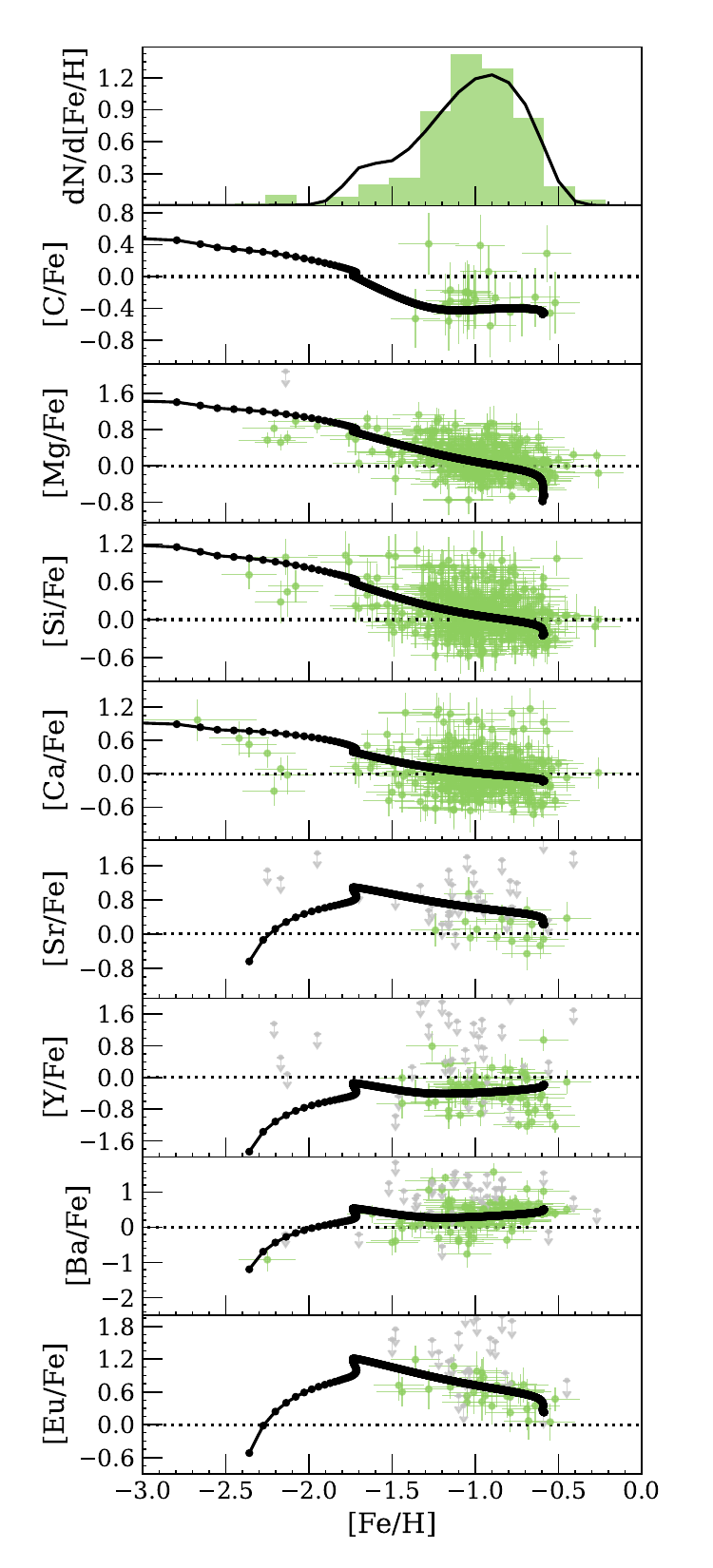}
    \caption{Metallicity distribution function (top) and [C/Fe], [Mg/Fe], [Si/Fe], [Ca/Fe], [Sr/Fe], [Y/Fe], [Ba/Fe], and [Eu/Fe] as a function of [Fe/H] for Fornax. Green points are abundance measurements, and gray arrows are upper limits. The black line shows the best-fit GCE model with NSMs as the only $r$-process source. This is the model that corresponds to the parameters in Table \ref{tab:fornax}.}
    \label{fig:Fornax_model}
\end{figure}

Figure \ref{fig:FornaxSFH} shows the SFH of Fornax from our model compared to one from Hubble Space Telescope photometry \citep{rusakov_bursty_2021}. Our model, which forms all of Fornax's stars in one 2 Gyr burst of star formation, does not reproduce the extended, bursty SFH indicated by photometry.


While the GCE model is overall poorly-suited for Fornax, we include it here in the Appendix because the neutron-capture abundances in Fornax still provide constraints to $t_{\rm min, NSM}$ in the model. When NSMs are the only $r$-process source, the best-fit model for Fornax has $t_{\rm min, NSM} = 11^{+2}_{-1}$ Myr. Though there are no [Eu/Fe] measurements at low ${\rm [Fe/H]} < -2$, the slope of the [Eu/Fe] distribution as a function of [Fe/H] must be able to place some constraints on the $r$-process delay time. The neutron-capture abundance measurements in Fornax corroborate our findings with Sculptor and Draco; namely that the $r$-process in dSphs must occur on a timescale similar to CCSNe.

\begin{deluxetable}{c|c}
\tablecolumns{2}
\tablewidth{0.5\linewidth}
\tablecaption{Best-fit parameters for Fornax \label{tab:fornax}}
\tablehead{
  \colhead{Parameter} & 
  \colhead{MCMC median} \\
}
\startdata
$A_{\rm in}$ & $1.28^{+0.19}_{-0.14}$ \\
$\tau_{\rm in}$ & $0.41^{+0.03}_{-0.03}$ \\
$A_{\rm out}$ & $1.41^{+0.04}_{-0.04}$ \\
$A_{*}$ & $9.81^{+0.14}_{-0.36}$ \\
$\alpha$ & $0.32^{+0.02}_{-0.02}$ \\
$A_{\rm NSM}$ & $8.95^{+0.67}_{-0.83}$ \\
$t_{\rm min,NSM}$ (Myr) & $11^{+2}_{-1}$ \\
${\rm normSr}_{\rm NSM}$ & $0.31^{+0.07}_{-0.05}$ \\
${\rm normY}_{\rm NSM}$ & $0.01^{+0.002}_{-0.002}$ \\
\hline
$-\ln{L}$ & 2337 \\
\enddata
\tablecomments{Values in this table follow the same format as Table \ref{tab:Iatests}.}
\end{deluxetable}

\section{MCMC corner plot} \label{appendix:corner}

\begin{figure}
    \centering
    \includegraphics[width=1.0\linewidth]{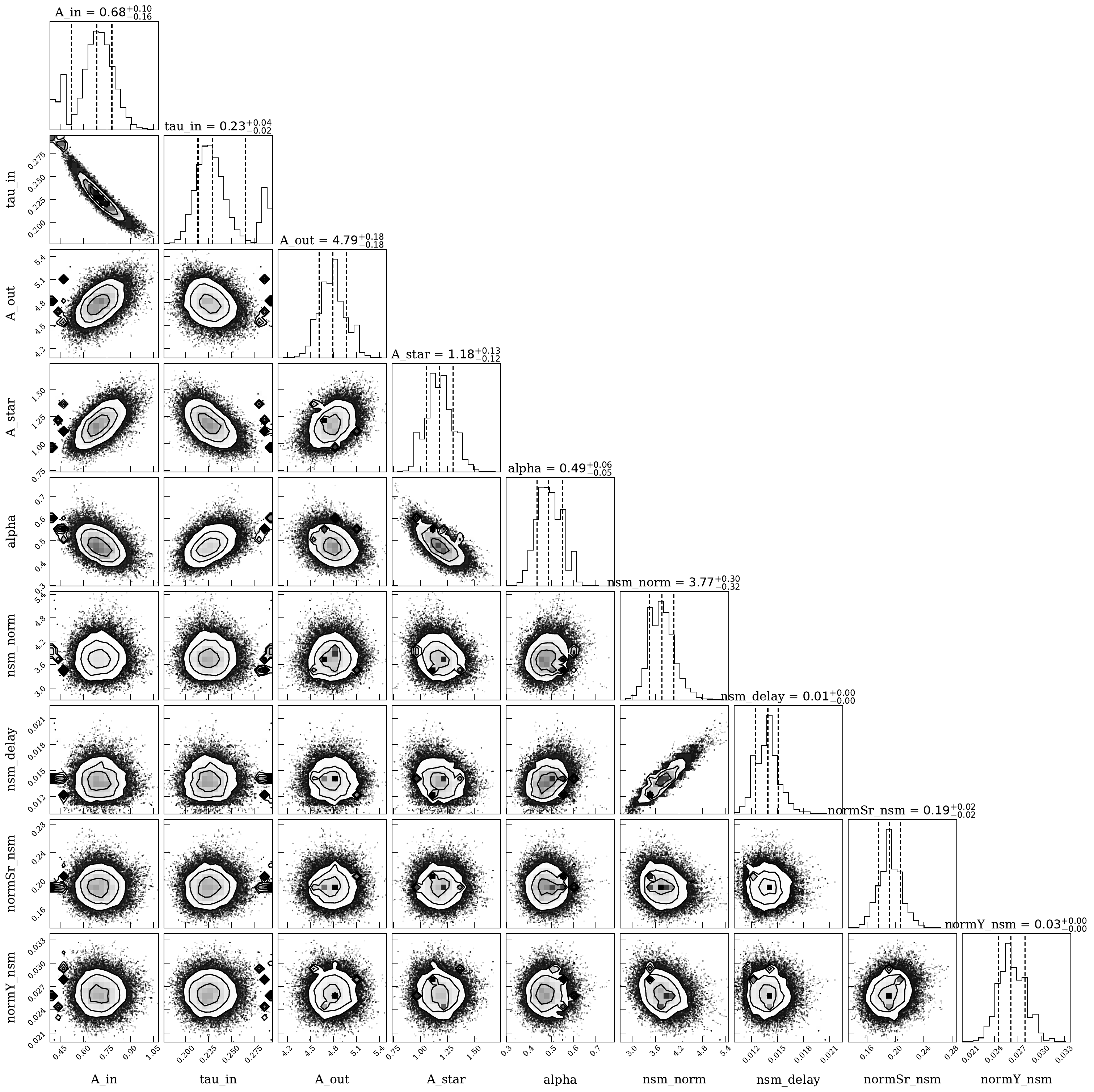}
    \caption{Corner plot from \texttt{emcee}. The parameters shown here are fit to the Sculptor dSph with NSMs as the only $r$-process source in the model. The parameters nsm\textunderscore norm and nsm\textunderscore delay correspond to $\rm A_{NSM}$ and $\rm t_{\rm min, NSM}$.}
    \label{fig:cornerplot}
\end{figure}

In Figure \ref{fig:cornerplot} we show the \texttt{emcee} corner plot from fitting the GCE model to Sculptor with NSMs as the only $r$-process source. The best-fit parameters from this MCMC fit are given in the first column in Table \ref{tab:r_tests}. The corner plot shows the posterior probability distributions of the free parameters. Two pairs of the GCE model parameters show clear correlations: $\rm A_{*}$ with $\alpha$ and $\rm A_{NSM}$ with $\rm t_{\rm min, NSM}$ (these are denoted as A\textunderscore star, alpha, nsm\textunderscore norm, and nsm\textunderscore delay in the figure). These correlations are not surprising. $\rm A_{*}$ and $\alpha$ both control star formation; when the normalization of the star formation law, $\rm A_{*}$, increases, the power-law index, $\alpha$, must decrease to control the star formation rate (see Table \ref{tab:params}). Likewise, when the minimum NSM delay time, $\rm t_{\rm min, NSM}$ , increases, the number of NSMs, $\rm A_{NSM}$, must increase as well to produce enough $r$-process material to match the observed abundances.

\end{appendix}

\bibliography{SMAUG_refs_ADS}{}
\bibliographystyle{aasjournalv7}



\end{document}